\documentclass[a4paper,11pt]{article}
\pdfoutput=1 

\usepackage{jcappub} 

\usepackage{soul}
\usepackage[T1]{fontenc} 
\usepackage{xcolor}
\usepackage[toc,page]{appendix}
\usepackage[normalem]{ulem}

\newcommand*{\blue}[1]{{\textcolor{blue}{#1}}}

\definecolor{grassgreen}{RGB}{86, 160, 50}

\definecolor{darkgreen}{rgb}{0.0, 0.5, 0.0}
\title{\boldmath Cosmic homogeneity: the effect of redshift-space distortions and bias and cosmological constraints}

\author[a,1]{Xiaoyun Shao,\note{Corresponding author}}
\author[b,a]{Rodrigo Gon\c{c}alves,}
\author[a]{Carlos A. P. Bengaly,}
\author[c]{Gabriela C. Carvalho,}
\author[a]{Jailson Alcaniz}

\affiliation[a]{Observat\'orio Nacional, Rio de Janeiro - RJ, 20921-400, Brasil}
\affiliation[b]{ Departamento de F\'{\i}sica, Universidade Federal Rural do Rio de Janeiro, Serop\'edica - RJ, 23897-000, Brasil}
\affiliation[c]{Faculdade de Tecnologia, Universidade do Estado do Rio de Janeiro, 27537-000, Resende, RJ, Brazil}

\emailAdd{xiaoyun48@on.br; rsousa@on.br; carlosbengaly@on.br; gabriela.coutinho@fat.uerj.br; alcaniz@on.br}

\abstract{We present a novel cosmological analysis based on the angular correlation dimension $D_2$ curve, a cumulative statistic derived from the two-point correlation function. Unlike traditional 3D approaches, angular $D_2$ is inherently less sensitive to nonlinear dynamical distortions, such as the small-scale Finger-of-God (FoG) effect. Using both MultiDark-Patchy and EZmock galaxy catalogs, we assess the scale-dependent impact of redshift-space distortions on $D_2$ and bias measurements. We demonstrate that the systematic errors associated with FoG modeling can be significantly reduced by restricting the analysis to appropriate minimum comoving angular scales of $\sim 1.25^{\circ} $, which corresponding to physical scales of $18$–$23\,h^{-1}\,\mathrm{Mpc}$ over the redshift range $0.46 \leq z \leq 0.74$ within the standard $\Lambda$CDM model. Since the observational estimative of $D_2(\theta)$ is not dependent on a cosmological model we obtain robust estimates of the galaxy bias and place competitive constraints on the physical matter density $\omega_m$. By applying this framework to SDSS DR12 and DR16 Luminous Red Galaxy data, we obtain $\omega_m = 0.137^{+0.041}_{-0.059}$ (1$\sigma$), which agrees with current CMB analyses. Our results highlight the potential of the angular $D_2$ curve as a model-independent and robust tool for cosmological parameter inference.}

\begin{document}
\maketitle
\flushbottom

\section{Introduction}\label{sec:intro}

The $\Lambda$-Cold Dark Matter ($\Lambda$CDM) framework forms the foundation of the Standard Cosmological Model (SCM), describing an expanding universe undergoing a late-time acceleration phase. Despite its success in explaining a wide range of cosmological observations - such as the Cosmic Microwave Background (CMB), the clustering and weak lensing of large-scale structures (LSS), and distances to standard candles like Type Ia Supernovae (SN)~\cite{aghanim2021planck,Brout:2022vxf,eBOSS:2020yzd,DES:2021wwk,ACT:2023kun,Li:2023tui}  - key components of the model, namely Cold Dark Matter (CDM) and dark energy ($\Lambda$), remain mysterious. 

The SCM faces significant theoretical and observational challenges. Theoretical issues include the coincidence and fine-tuning problems~\cite{Weinberg:2000yb}, while observationally, a roughly 5$\sigma$ tension exists between early and late-time measurements of the Hubble constant (see, e.g. \cite{DiValentino:2021izs} and references therein). Moreover, recent observations from the Dark Energy Spectroscopic Instrument (DESI) have hinted at the possibility of dynamic dark energy instead of a constant $\Lambda$~\cite{DESI:2024mwx}, although this result remains debated~\cite{DESI:2024aqx,Cortes:2024lgw}. These challenges emphasize the importance of rigorously testing the fundamental assumptions of the SCM as any significant deviation would require a complete reformulation of the cosmological paradigm.

In view of these ongoing challenges, it becomes imperative to rigorously examine the foundational assumptions underlying the $\Lambda$CDM model. A central tenet of this framework is the Cosmological Principle (CP), which posits that the Universe is statistically homogeneous and isotropic on sufficiently large scales~\cite{clarkson2010inhomogeneity,Maartens:2011yx,Clarkson:2012bg,Aluri:2023dsf}.

During the 1970s, Benoit Mandelbrot \cite{mandelbrot1982fractal,mandelbrot1997galaxy} showed how the fractal or Hausdorff dimension (introduced in 1918 by mathematician Felix Hausdorff) is a rigorous way to characterize geometrical irregularities and chaos in Nature. The Hausdorff dimension estimate is independent of the amplitude of such irregularities and it only depends on how they scale with size. These ideas have been applied to galaxy surveys as a way to test if the metric of the Universe is geometrically homogeneous and isotropic~\cite{Hogg:2004vw,scrimgeour2012wigglez,alonso2014measuring,alonso2015homogeneity,laurent201614,ntelis2017exploring,gonccalves2018cosmic,gonccalves2018measuring,HomogeneityQuasars,bengaly2018probing,gonccalves2021measuring,andrade2022angular,mittal2024cosmic,Shao:2023sxk,Shao:2024qrd}, as well as to the CMB \cite{camacho2022measurement,Shao:2025djg}.

An important way to investigate the CP is the analysis of the cosmic homogeneity through the fractal correlation dimension $D_2$~\cite{scrimgeour2012wigglez}, specifically taking into account the advantages of being intrinsically more sensitive to the shape of the large-scale structure clustering patterns than their amplitude. In the literature, there are two main approaches to the use of $D_2$. The first one is based on the spatial 3-dimensional separation $r$ (so $D_2 = D_2(r)$) of matter tracers, such as galaxies~\cite{gonccalves2018cosmic} and quasars~\cite{gonccalves2018measuring,gonccalves2021measuring}. However, this analysis is model-dependent, since we need to assume a cosmological model to convert redshifts into distances, and thus it can only provide a consistency test of the CP assumption. The second approach is based on the two-dimensional angular distribution of points, characterized by the angular separation $\theta$ (i.e., $D_2 = D_2(\theta)$). Conversely from the former approach, this one can be carried out in a model independent way, because there is no conversion from redshift to distances. 
In both cases, $D_2$ determine the scale on which the universe can be defined as homogeneous, there is the so called homogeneity scale ($r_h$) and angular homogeneity scale ($\theta_h$).
 
The homogeneity scale $r_h$ has been identified in the range $\sim 70$–$150\,h^{-1}\mathrm{Mpc}$ using data from several large-scale galaxy surveys \cite{laurent201614,HomogeneityQuasars}. As proposed by \cite{gonccalves2018cosmic}, its angular counterpart, $\theta_h$, exhibits a clear redshift dependence, with the transition angular scale increasing toward lower redshifts, in agreement with the expectations of the standard cosmological paradigm. This behavior is expected, since matter perturbations grow stronger at later epochs, making the Universe appear progressively clumpier as the redshift decreases.

Moreover the cosmic homogeneity can be related with the cosmic composition. More specifically the cosmological interpretation of the matter homogeneity scale is consistent with the dependence of cosmological parameters when we assume a fiducial cosmology. Given this dependence, it is of interest to explore whether the homogeneity scale can be used to constrain cosmology. Indeed, several studies have already derived cosmological constraints from the homogeneity scale \cite{Ntelis:2018ctq,Shao:2024qrd} and references therein.

In this paper, our main objective is to extend previous analyses that used the homogeneity scale to constrain cosmology by considering a broader portion of the angular homogeneity index $D_2(\theta)$.  While homogeneity is only recovered asymptotically on large scales, the $\Lambda$CDM framework successfully describes structure formation on scales below the homogeneity scale, apart from deeply nonlinear regimes where modeling becomes challenging. Therefore, it is important to determine an appropriate minimum scale threshold above which the model remains valid, and thus can be reliably used to derive cosmological constraints. In this context, we investigate how the angular homogeneity index $D_2(\theta)$ can be used to  constrain cosmological parameters.
Note that this approach exhibits a reduced bin-to-bin correlations compared to the traditional two-point correlation function (2PCF)~\cite{andrade2022angular,Shao:2023sxk,Shao:2024qrd}.

For instance, in redshift space, the clustering of matter tracers exhibits distortions due to their peculiar velocities, which serves as a powerful probe for testing theories of gravity on large scales. This effect is called redshift-space distortion (RSD). There are two main aspects of those distortions, depending on the scale studied. On large scales, galaxies are falling into large gravitational potentials, which tend to sharpen their distribution along the line-of-sight in redshift space, a result that is known as Kaiser effect \cite{kaiser1987clustering}. On small scales, the random velocity dispersion of galaxies along the line of sight leads to the so-called Finger-of-God (FoG) effect~\cite{ballinger1996measuring}.

Early RSD measurements were primarily used to estimate the mean matter density, $\Omega_m$, which can be directly inferred from the growth rate $f(z)$ under a given gravity model \cite{beutler20116df,samushia2012interpreting}. Subsequently, it was demonstrated that RSD can also be used effectively to discriminate between dark energy and modified gravity scenarios \cite{chuang2013modelling,chuang2013clustering}. As a result, RSD gained prominence as a key tool for probing gravity, leading to widespread applications across galaxy redshift surveys, both in the local Universe and at higher redshifts up to $z \sim 1$. Notable examples include 6dFGS \cite{beutler20116df}, SDSS \cite{samushia2012interpreting,chuang2013modelling,chuang2013clustering}, WiggleZ \cite{blake2012wigglez,contreras2013wigglez}, VIPERS \cite{de2013vimos}, and BOSS \cite{tojeiro2012clustering,reid2012clustering,reid20142}.
However, extracting unbiased cosmological constraints from RSD data requires precise modeling of the clustering signal, particularly at small scales where nonlinear dynamics become significant. This necessitates a careful characterization of the connection between observed tracers and the underlying dark matter distribution. However, discrepancies between observational measurements and the predictions of $\Lambda$CDM continue to persist, even when sophisticated RSD models are employed \cite{macaulay2013lower,marulli2017redshift}.

Taking into account the advantages of the fractal dimension, we investigate how systematic errors that arise from RSD effects  depend on the angular comoving scale and its impact on the estimation of both the bias parameter and cosmological parameters. While much of the existing literature has focused on determining the angular homogeneity scale ($\theta_H$) \cite{alonso2014measuring,gonccalves2018cosmic}, the present work instead concentrates on assessing the impact of nonlinear effects on the full angular homogeneity index $D_2$. To achieve this, we first make use of realistic mock samples, specifically the MultiDark-Patchy mock catalogs \cite{kitaura2016clustering, rodriguez2016clustering} and the EZmock galaxy catalogs \cite{zhao2021completed}, which are designed to replicate the SDSS DR12 and DR16 Luminous Red Galaxy (LRG) samples. We assess the systematic errors on the bias factor using these mock catalogs, and further extend the analysis to investigate systematic errors on the physical matter density $\left( \omega_m \right)$ using real data from the SDSS DR12 and DR16 catalogs.  We demonstrate that these systematic errors can be substantially reduced by restricting the fit to an appropriate range of comoving scales.

This paper is organized as follows: Sec.~\ref{sec:Dataset} describes the dataset used in this work. Sec.~\ref{sec:method} presents the methodology, including the theoretical framework and the estimation of the angular correlation dimension $D_2(\theta)$.
Sec.~\ref{sec:result1} investigates the impact of RSD on $D_2$.
Sec.~\ref{sec:s5} presents the analysis of the linear bias factor using mock catalogs and SDSS DR12 and DR16 data.
Sec.~\ref{sec:result} presents the cosmological constraints on $\omega_m$. We summarize our conclusions in  Sec.~\ref{sec:conclu}.

\section{Data}\label{sec:Dataset}

The Sloan Digital Sky Survey is a global scientific collaboration that has produced highly accurate three-dimensional maps of the Universe. The project was partitioned into four distinct phases: SDSS-I (2000-2005), SDSS-II (2005-2008), SDSS-III (2008-2014), and SDSS-IV (2014-2020). The present study focuses on the Data Release 12 (DR12) of the Baryon Oscillation Spectroscopic Survey\footnote{https://live-sdss4org-dr12.pantheonsite.io/} (BOSS)~\cite{dr12a,dr12b,dr12c}, and the Data Release 16 (DR16) of the extended Baryon Oscillation Spectroscopic Survey\footnote{https://www.sdss4.org/dr16/} (eBOSS)~\cite{dr16}, which are subsets of the Sloan Digital Sky Survey III (SDSS-III) and Sloan Digital Sky Survey IV (SDSS-IV), respectively.

\begin{figure*}[!b]
	\centering
    \includegraphics[width=0.6\textwidth]{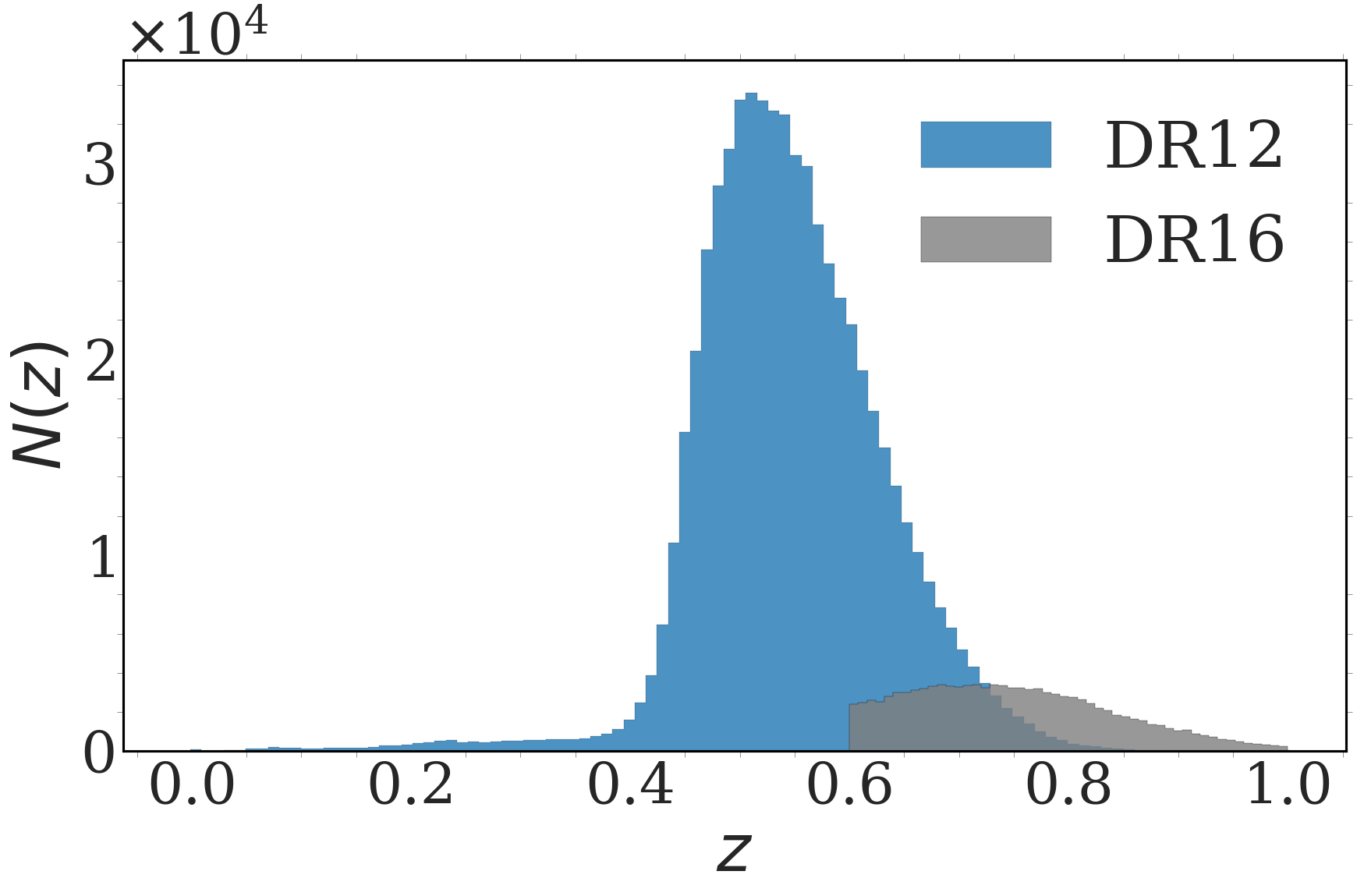} 
    \caption{Redshift distributions of the LRGs from BOSS DR12 and eBOSS DR16 catalogues. We considered the redshift range $0.46 < z < 0.74$.}
\label{fig:hist}
\end{figure*}

\begin{figure*}[!t]
	\centering
    \includegraphics[width=0.45\textwidth]{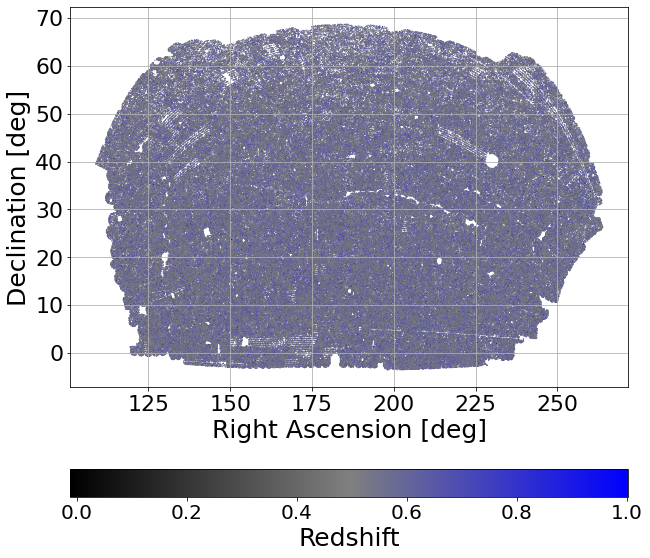} 
    \includegraphics[width=0.45\textwidth]{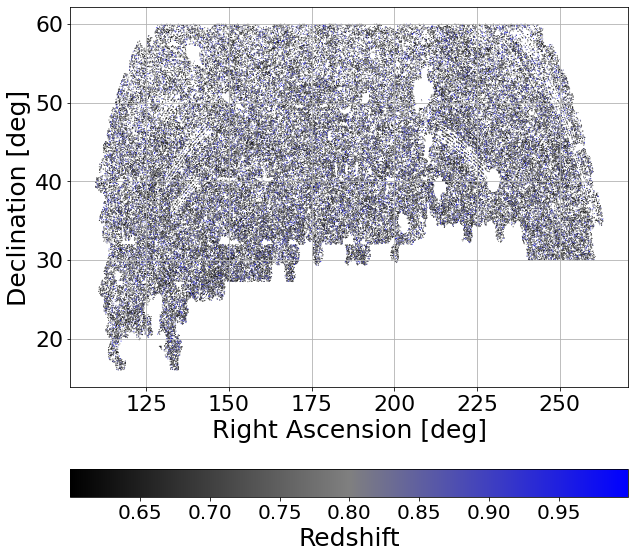} 
    \caption{The footprint of DR12 and DR16, respectively}
\label{fig:dr16sky}
\end{figure*}

Specifically this work adopts the northern sky of both BOSS DR12 and eBOSS DR16 catalogues of Luminous Red Galaxies and for the sake of simplicity we hereafter name them as DR12 and DR16, respectively. The exclusion of the southern sky subsample is due to the limited sky coverage. The main features of the DR12 and DR16 data set used in this work are shown in Table~\ref{t10} and Table~\ref{t20}, respectively. The redshift interval in each dataset is $0.46 < z < 0.62$ with around 420,000 points (DR12) and $0.67 < z < 0.74$ with approximately 30,000 points (DR16), as displayed in Fig.\ref{fig:hist}. The footprint of the sky area coverage of both catalogues is shown in Fig.\ref{fig:dr16sky}.

We split the data into redshift bins of $\Delta z = 0.01$, with the size of the bin being chosen to avoid projection effects \cite{alonso2014measuring}, as discussed in Appendix~\ref{bin7}. This choice allows us to avoid possible biases on homogeneity scale measurements, and increase the number of data points, thus providing a good statistical performance for the analysis \cite{Goncalves:2018sxa, Goncalves:2020erb, Alonso:2013boa, Andrade:2022imy}.

To test the robustness of the method and compute the covariance matrix, we rely on 1000 mock catalogs from the MultiDark-Patchy mock catalogs \cite{kitaura2016clustering, rodriguez2016clustering} and the EZmock galaxy catalogues \cite{zhao2021completed}.

\begin{table}[!htbp]\centering 
\caption{The redshift bins adopted in our analysis from BOSS (DR12), along with the redshit bin means and corresponding number of LRGs in each bin.}
\label{t10}
\begin{tabular}{cccc}
\hline 
$z$ ~~&~~ $\bar{z}$ ~~&~~ $N_{gal}$ ~~\\  \hline ~~ 0.46-0.47 ~~&~~ 0.465 ~~&~~ 22551 ~~\\~~  0.47-0.48 ~~&~~ 0.475 ~~&~~ 27319 ~~\\~~  0.48-0.49 ~~&~~ 0.485 ~~&~~ 29251 ~~\\~~  0.49-0.50 ~~&~~ 0.495 ~~&~~ 31763 ~~\\~~  0.50-0.51 ~~&~~ 0.505 ~~&~~ 33107 ~~\\~~  0.51-0.52 ~~&~~ 0.515 ~~&~~ 32887 ~~\\~~  0.52-0.53 ~~&~~ 0.525 ~~&~~ 32794 ~~\\~~  0.53-0.54 ~~&~~ 0.535 ~~&~~ 31995 ~~\\~~  0.54-0.55 ~~&~~ 0.545 ~~&~~ 31355 ~~\\~~  0.55-0.56 ~~&~~ 0.555 ~~&~~ 29486 ~~\\~~  0.56-0.57 ~~&~~ 0.565 ~~&~~ 28995 ~~\\~~  0.57-0.58 ~~&~~ 0.575 ~~&~~ 25289 ~~\\~~  0.58-0.59 ~~&~~ 0.585 ~~&~~ 23997 ~~\\~~  0.59-0.60 ~~&~~ 0.595 ~~&~~ 22568 ~~\\~~  0.60-0.61 ~~&~~ 0.605 ~~&~~ 20594 ~~\\~~  0.61-0.62 ~~&~~ 0.615 ~~&~~ 18799 ~~\\
\hline 
\end{tabular} 
\end{table}

\begin{table}[!htbp]\centering \caption{The redshift bins adopted in our analysis from eBOSS (DR16), along with the redshit bin means and corresponding number of LRGs in each bin.}
\label{t20}
\begin{tabular}{cccc}
\hline 
$z$ ~~&~~ $\bar{ z}$ ~~&~~ $N_{gal}$ ~~\\ \hline ~~ 0.67-0.68 ~~&~~ 0.675 ~~&~~ 4073 ~~\\~~  0.68-0.69 ~~&~~ 0.685 ~~&~~ 4222 ~~\\~~  0.69-0.70 ~~&~~ 0.695 ~~&~~ 4064 ~~\\~~  0.70-0.71 ~~&~~ 0.705 ~~&~~ 4209 ~~\\~~  0.71-0.72 ~~&~~ 0.715 ~~&~~ 4239 ~~\\~~  0.72-0.73 ~~&~~ 0.725 ~~&~~ 4084 ~~\\~~  0.73-0.74 ~~&~~ 0.735 ~~&~~ 4210 ~~\\
\hline 
\end{tabular} 
\end{table}

\newpage


To correct for known clustering systematics, a specific weight is applied to each galaxy. For the DR12 samples, we followed the weighting scheme described in \cite{reid2016sdss,Ntelis:2017nrj}, where the weight assigned to each galaxy is given by
\begin{equation}
    w_{\text {gal }}=w_{\mathrm{FKP}} * w_{\mathrm{systot}} *\left(w_{\mathrm{cp}}+w_{\mathrm{noz}}-1\right)
\end{equation}
where we use the FKP weight, $w_{\mathrm{FKP}}$, \cite{feldman1993power} in
order to reduce the variance of the two-point correlation function estimator. $w_{\text {systot }}=w_{\text {star }} * w_{\text {see }}$ is the total angular systematic weight accounting for the seeing effect and the star confusion effect;
$w_{\text{cp}}$ accounts for the fact that the survey cannot spectroscopically observe two objects that are closer than $62^{\prime \prime}$~\cite{reid2016sdss,Ntelis:2017nrj}, and $w_{\text{noz}}$ accounts for redshift failures.

For the DR16 samples, we adopted the weighting described in~\cite{ross2020completed,bautista2021completed,nadathur2020completed}. Each galaxy is weighted as:
\begin{equation}
w_{\mathrm{tot}}=w_{\mathrm{noz}} w_{\mathrm{cp}} w_{\mathrm{sys}} w_{\mathrm{FKP}}
\end{equation}
where $w_{\text{sys}}$ is applied to each object to correct for imaging systematics in DR16.

\section{Correlation dimension and angular homogeneity scale}\label{sec:method}

\subsection{Theoretical framework}

In this section, we make an overview of the method that was developed to assess the correlation dimension ($D_{2}$). For a complete description, we refer the reader to~\cite{Shao:2023sxk,Shao:2024qrd}, where this method is fully described.

The correlation dimension \(D_2\) is the key quantity for measuring the angular homogeneity scale of the data (see, e.g., \cite{Hogg:2004vw, Sarkar:2009iga, Scrimgeour:2012wt, Pandey:2013xz, Pandey:2015xea, Sarkar:2016fir, Laurent:2016eqo, Ntelis:2017nrj, Goncalves:2018sxa, Goncalves:2020erb, Kim:2021osl, Labini:2009ke, Labini:2011dv, Park:2016xfp, Heinesen:2020wai, Alonso:2013boa, Alonso:2014xca, Goncalves:2017dzs, Andrade:2022imy, Ntelis:2018ctq, Ntelis:2019rhj, Nesseris:2019mlr, Shao:2023sxk} and references therein). In a few words the angular homogeneity scale ($\theta_H$) is reached when $D_2(\theta_H) = 1.98$, by assuming an observational threshold of $1\%$~\blue{\cite{scrimgeour2012wigglez}.}

As we are conducting a two-dimensional (2D) analysis, the correlation dimension is defined over the angular separation \(\theta\). In this case the expression for the correlation dimension is given by
\begin{eqnarray}
\label{eq:d2xi}
D_{2}(\theta)=2+\frac{d \ln}{d \ln \theta}\left[1+\frac{1}{1-\cos \theta} \int_{0}^{\theta} \omega\left(\theta^{\prime}\right) \sin \theta^{\prime} d \theta^{\prime}\right],
\label{w2}
\end{eqnarray}
where $\omega(\theta)$ is the 2D two-point angular correlation function (2PACF), which can be expressed as~\cite{crocce2011modelling}
\begin{eqnarray}
\label{eq:omega}
\omega(\theta)=\int d z_{1} W\left(z_{1}\right) \int d z_{2} W\left(z_{2}\right) \xi\left(r\left(z_{1}\right), r\left(z_{2}\right), \theta, \bar{z}\right)\label{eq:d2w1}
\end{eqnarray}
where $\xi$ is the 3D two-point correlation function for matter, as described below. Also, we assume $W(z) \equiv b(z) \phi(z)$, where $\phi(z)$ denotes the radial selection, i.e., the probability to include a galaxy in a given redshift bin, and $b(z)$ is the bias parameter, which relates the galaxy distribution to the underlying matter distribution. 

In this paper, we consider only top-hat window functions and narrow redshift bins, so we compute the 2PACF and the 3D correlation function for matter as
\begin{equation}
\xi(r)=\frac{1}{2\pi^2}\int_0^\infty j_0(k r) k^2 P(k) dk,\label{eq:corfunc}
\end{equation}
where $P(k)$ is the matter power spectrum and $j_0(x)=\sin{(x)}/x$. Also, the comoving distance to a certain redshift $z$ is given by 
\begin{equation}
r(z)=\int_{0}^{z} \frac{c}{H(z')} dz',
\label{eq:rz}
\end{equation}
where  
\begin{equation}
H(z)=H_0\sqrt{\Omega_{m0}(1+z)^{3} + (1-\Omega_{m0})} \; ,
\label{eq:Hz}
\end{equation}
where \(H_0\) is the present-day Hubble parameter.

\subsection{Observational estimates of $D_{2}$ and $\theta_H$}\label{ObsEst}

\noindent

The above method describes how to obtain the theoretical predictions of $D_{2}(\theta)$. In what follows, we discuss how to estimate the relevant observational quantities. We assume the Landy-Szalay 2PACF estimator, $\hat{\omega}_{ls}(\theta)$~\cite{landy1993bias}, which is obtained directly from a combination of the data and random catalogues~\cite{Alonso:2013boa,Goncalves:2017dzs,Andrade:2022imy}. It is defined as
\begin{equation}
\label{eq:w_ls}
\widehat{\omega}_{ls}(\theta) = \frac{DD(\theta) -2DR(\theta) + RR(\theta)}{RR(\theta)},
\end{equation}
where $DD(\theta)$, $DR(\theta)$, and $RR(\theta)$ are the numbers of pairs of data as a function of there separations $\theta$, normalized by the total number of pairs, in data-data, data-random and random-random catalogues, respectively. The pair-counting was done with the {\sc TreeCorr} package~\cite{jarvis2004skewness}. Concerning to the random catalogues, we use the BOSS and eBOSS catalogues, which are approximately 50 and 20 times larger than the respective real catalogues~\cite{dr12a,dr12b,dr12c,dr16}. This ensures that the statistical fluctuations caused by the random points are insignificant.

Therefore, in order to calculate the $D_2$ values and uncertainties, we perform the following steps: (i) we calculate the value of $D_2$ using Eqs.~\ref{eq:d2xi} and \ref{eq:w_ls} for different values of $\theta$; (ii) we repeat this procedure 1000 times via bootstrap resampling technique; (iii) we calculate the mean and standard deviation of the values obtained in step (ii) as our measurements of the $D_2$ and its corresponding uncertainty.

\begin{figure}[!t]
    \centering
    \includegraphics[width=0.49\textwidth]{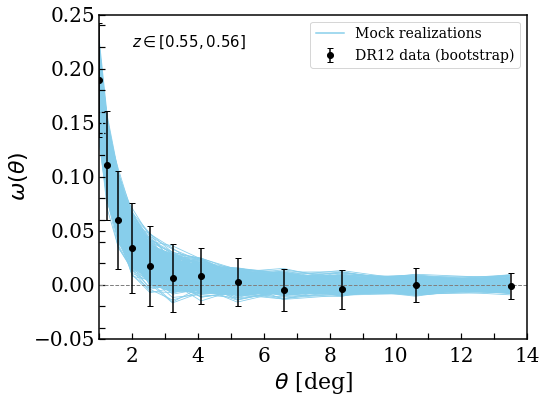} 
    \includegraphics[width=0.49\textwidth]{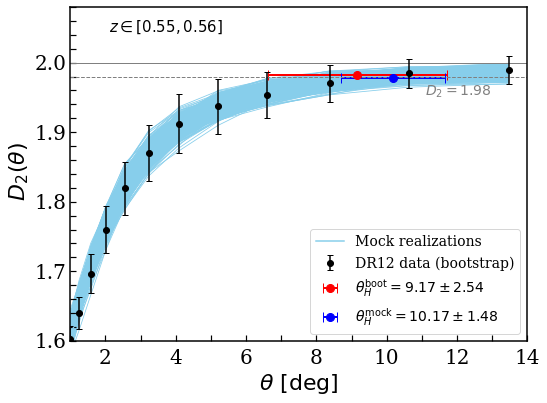} 
    \caption{Observed angular correlation function (left) and angular correlation dimension $D_2 (\theta)$ (right) for DR12 in the Redshift Bin $z \in [0.55,0.56]$} 
\label{fig:d2}
\end{figure}

We perform this analysis by using the observational samples mentioned in the previous section (SDSS DR12 and DR16, Tables~\ref{t10} and~\ref{t20}, respectively) for each redshift bin. For the sake of completeness Table~\ref{theta_all} (Appendix~\ref{ThetaHBinsZ}) presents the angular homogeneity scales $\theta_H$ across all redshift bins, including results from both bootstrap resampling and mock catalogs. This table allows a direct comparison between the observational and simulated estimates of the homogeneity scale, and provides insight into the consistency and redshift evolution of large-scale structure homogeneity. For the sake of example, we present the result for the bin $0.55 < z < 0.56$. Fig.~\ref{fig:d2} presents the observed angular correlation function $\omega(\theta)$ (left)  and the angular correlation dimension  $D_2 (\theta)$ (right) as a function of angular separation $\theta$ (in degrees) in the redshift bin $z \in [0.55,0.56]$. Black points show the measured values from the data, with error bars estimated via 1000 bootstrap resampling. Light blue curves correspond to 1000 mock catalogs, illustrating the expected range of $D_2 (\theta)$ under $\Lambda$CDM model. The horizontal dashed line at $D_2(\theta) = 1.98$ defines the homogeneity threshold.
Red and blue error bars indicate the homogeneity scale $\theta_{H}$, defined as the scale at which $D_2(\theta)$ approaches 1.98, estimated from bootstrap (red) and mocks (blue), respectively, $\theta_{H}^{\text {boot }}=9.17 \pm 2.54^{\circ}$, and  $\theta_{H}^{\text {mock }}=10.75 \pm 1.53^{\circ}$.
The comparison shows that both data and mocks reach homogeneity on mean scales around
$9^\circ$-$10^\circ$, at the redshift mean $\bar{z} = 0.555$, consistent with $\Lambda$CDM expectations.

\begin{figure*}[!t]
    \centering
    \includegraphics[width=0.4\textwidth]{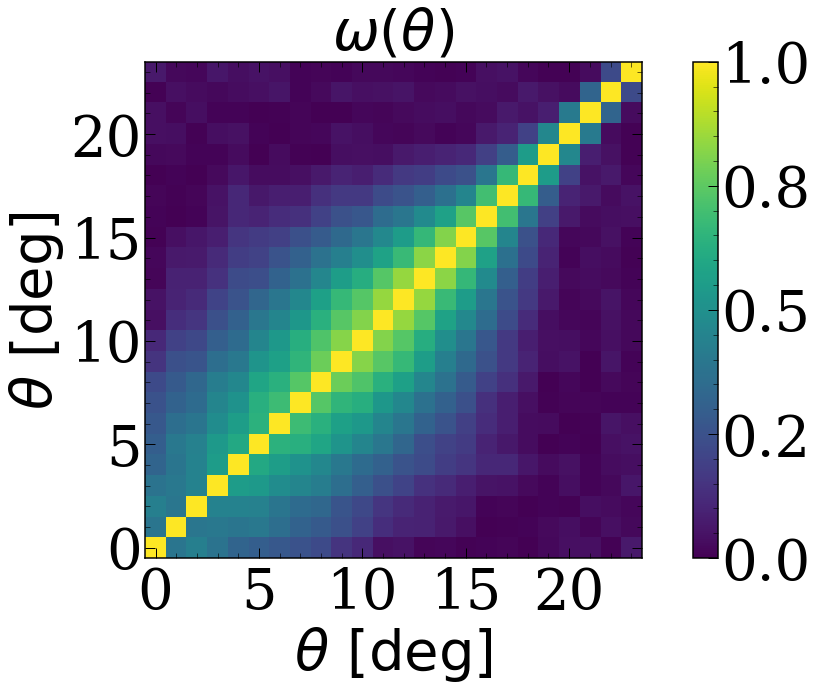} 
    \includegraphics[width=0.4\textwidth]{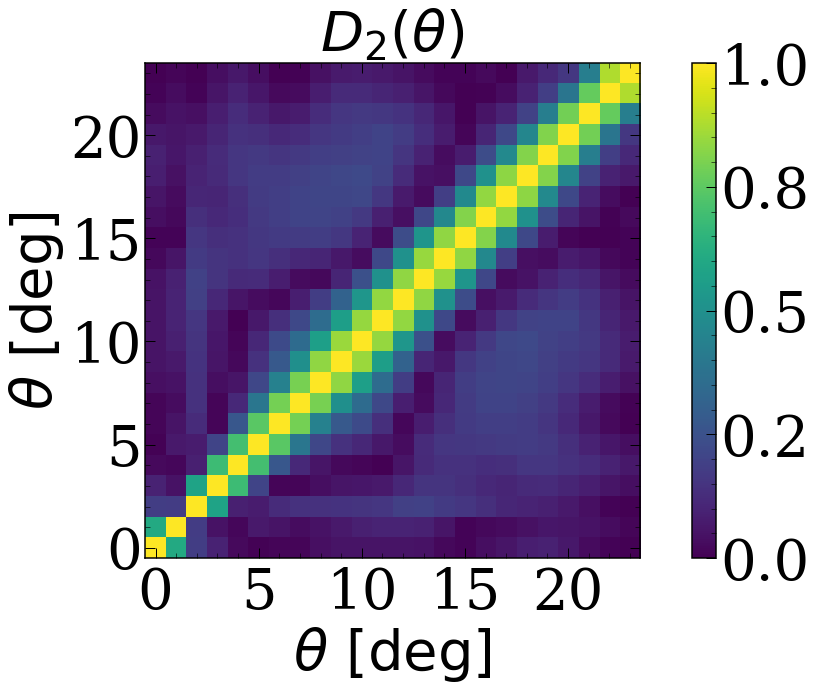} 
    \caption{Absolute Correlation Matrices $|\rho_{ij}|$ of $\omega(\theta)$ and $D_2(\theta)$ from 1000 Mock Realizations in the Redshift Bin $z \in [0.55, 0.56]$} 
\label{fig:cov}
\end{figure*}

Fig.~\ref{fig:cov} shows the absolute Pearson correlation matrices $|\rho_{ij}|$ computed from 1000 mock realizations of the angular correlation function $\omega(\theta)$ (left) and the angular correlation dimension $D_2(\theta)$ (right). Each pixel represents the strength of the correlation between two angular bins $\theta_i$ and $\theta_j$, with $|\rho_{ij}| = 1$ (yellow) indicating perfect correlation and $|\rho_{ij}| = 0$ (dark purple) indicating no linear correlation. The $D_2(\theta)$ matrix exhibits significantly reduced off-diagonal correlations compared to $\omega(\theta)$, indicating lower bin-to-bin covariance. This highlights the advantage of $D_2(\theta)$ as a decorrelated and robust summary statistic for angular clustering analysis.

\section{Theoretical impact of RSD on angular homogeneity index $D_2$}\label{sec:result1}

\subsection{Redshift-Space Distortions}
\label{rsd}

\noindent

A deeper study on the two-point correlation function should take into account non-linear effects on the matter power spectrum.  The theoretical prediction of the non-linear \(P(k)\) is commonly computed by using only {\sc CAMB}~\cite{lewis2000efficient} in combination with HALOFIT corrections~\cite{smith2003stable}. However, it is necessary to model the impact of RSD on these theoretical predictions.

With effect, in real space, the matter power spectrum, denoted by \(P_{\delta\delta}^{(r)}(k)\), describes the clustering of the underlying dark matter density field.  To model the galaxy power spectrum in redshift space, \(P_{gg}^{(s)}(k)\), two main effects must be taken into account. On large scales, the Kaiser effect leads to a modification on the power spectrum given by

\begin{equation}
    P_{gg}^{(s)}(k, \mu; b) = b^2 \left(1 + \beta \mu^2\right)^2 P_{\delta\delta}^{(r)}(k),
    \label{ka}
\end{equation}
where \(b\) is the linear bias factor, \(\beta = f/b\), and \(f\) is the linear growth rate, approximated as \(f \simeq \Omega_m^{0.55}(z)\)~\cite{Linder:2007hg}. 
On small scales, the distortion originated from the Finger-of-God is a non-linear effect that can be modeled using a Gaussian damping function that depends on the pairwise velocity dispersion \(\sigma_p\)~\cite{ballinger1996measuring}:

\begin{equation}
    \ln D(k, \mu; \sigma_p) = -\frac{1}{2} \left( \frac{k \sigma_p \mu}{H_0} \right)^2 \; .
    \label{fin}
\end{equation}
Combining both effects, the full galaxy power spectrum in redshift space is expressed as:
\begin{equation}
    P_{gg}^{(s)}(k; b, \sigma_p) = b^2 \int_0^1 \left(1 + \beta \mu^2\right)^2 D(k, \mu; \sigma_p) \, d\mu \, P_{\delta\delta}^{(r)}(k).
    \label{pk2}
\end{equation}
By applying a Fast Fourier Transform (FFT) to \(P_{gg}^{(s)}(k)\), we obtain the 3D two-point correlation function in redshift space:
\begin{equation}
    \xi^{(s)}(r; b, \sigma_p) = \mathrm{FFT}\left[ P_{gg}^{(s)}(k; b, \sigma_p) \right].
    \label{eq:fft}
\end{equation}

RSD plays a crucial role in the modelling of the power spectrum \(P(k)\) and, consequently, the angular homogeneity index \(D_{2}(\theta)\) (Eqs.~\ref{eq:d2xi} and~\ref{eq:omega}). 
We will explore the impact of RSD effects on the angular homogeneity index $D_{2}(\theta)$. On large scales, the Kaiser effect can be accurately modeled using Eq.~\ref{ka}, while on small scales, the FoG effect is typically modeled by Eq.~\ref{fin}. However, the theoretical model for RSD (Eq.~\ref{fin}) is not perfectly accurate at small scales due to the inherently nonlinear behavior of gravity in these regimes. One direct way to reduce the resulting systematic errors would be to improve the modeling of RSD at nonlinear scales (see, e.g., \cite{scoccimarro2004redshift,taruya2010baryon,seljak2011distribution,wang2014cdnet,de2013vimos}). In this work, we will adopt a different strategy, as described in Section.~\ref{sec:s5}.

\subsection{$D_2$ as a function of peculiar velocity dispersion}

\noindent

\begin{figure*}[!t]
    \centering
    \includegraphics[width=0.8\textwidth]{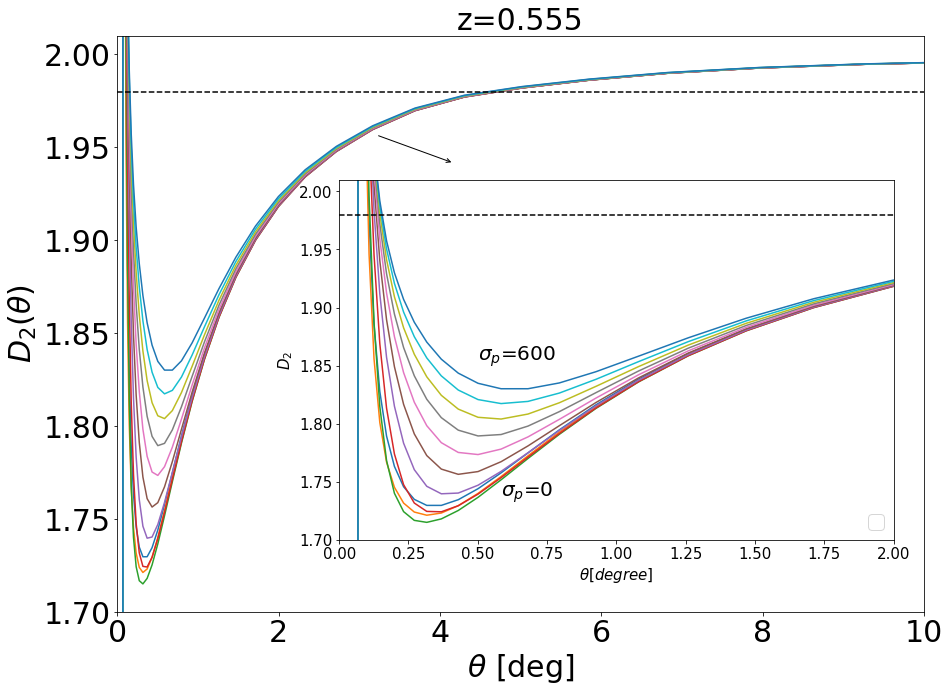} 
    \caption{Theoretical angular Correlation Dimension $D_2 (\theta)$ at $z=0.555$ for peculiar-velocity dispersion $\sigma_{p}$ $\in$ [0,600] km/s.} 
\label{fig:d2_th}
\end{figure*}

For the sake of illustration, Figure.~\ref{fig:d2_th} shows the theoretical angular correlation dimension $D_2 (\theta)$ as a function of angular separation $\theta$ (in degrees) at redshift  $z = 0.555$, accounting for the effects of line-of-sight velocity dispersion from $0$ to $600$ km/s. Each curve corresponds to a different value of $\sigma_p$, with $\sigma_p=0$ at the bottom and increasing upward to $\sigma_p=600$  km/s. Larger peculiar velocity dispersions increase the correlation dimension $D_2 (\theta)$ at small angular scales, driving it closer to the homogeneous value of $2$. The horizontal dashed line marks the homogeneity threshold $D_2 = 1.98$. The inset panel zooms in on small angular separations $\theta \in [0, 2^\circ] $.

According to \cite{Ntelis:2017nrj}, pairwise peculiar-velocity dispersion $\sigma_{p}$ for SDSS luminosity galaxies is around $250$ km/s. The range of velocity dispersion from $0$ to $600$ km/s is reasonable and well includes
the velocity dispersion of SDSS luminosity galaxies. This analysis shows that peculiar velocities have a significant scale-dependent impact on $D_2 (\theta)$ on small-degree angular separations. However, at large angular scales, the effect of peculiar velocity dispersions on $D_2(\theta)$ decreases significantly. We also carried out the same test in different redshift bins, and we found that the same conclusion is valid.

In the previous parts we explored how the theoretical angular homogeneity index $D_2$ depends on  the  velocity dispersion parameter $\sigma_P$.
In the next sections, we will investigate how these nonlinear effects affect the inferred galaxy bias and cosmological parameters.

\section{Linear bias factor}
\label{sec:s5}
In this section, we investigate how the systematic errors induced by nonlinear FoG effects depend on the choice of angular separation considered in the analysis. Specifically, we assess the accuracy of the theoretical model without the FoG effect as a function of \(\theta\). We verify that selecting an appropriate minimum angular scale cut for the fit of the whole $D_2(\theta)$ curve allows the impact of the FoG effect to be minimized, thereby making the contribution of nonlinear effects negligible. It is worth to note that in all analyses we are still taking into account the Kaiser effect. A key advantage of this approach is that it requires only a minimal number of free parameters for the modeling, simplifying the analysis while maintaining robustness.

\subsection{The FoG effect on bias for mock catalogs}

\noindent

The most direct way to investigate systematic errors from the non-linear FoG model is to use large numerical simulations and apply to them the same analysis techniques employed for real data. So we begin by infering  $D_{2}(\theta)$ using the MultiDark-Patchy mock catalogs and fitting the results with both the linear Kaiser model (Eq.~\ref{ka}, LIN model) and the nonlinear Finger-of-God model (Eq.~\ref{pk2}, NL model). We select subsets beginning at different minimum angular scales, $\theta_{\rm min}$\footnote{When we refer to a minimum angle we are implicitly describing an analysis through a whole range of angles, where $\theta_{min}$ represents the initial value.}, to be used as data for fitting. We perform fits using the LIN model and NL model over two ranges: $\theta \ge 0.5^\circ$ (so $\theta_{min} = 0.5^\circ$) and $\theta \ge 1.25^\circ$ (so $\theta_{min} = 1.25^\circ$). In these fits, the bias is treated as a free parameter and obtained by using the covariance matrix for the mock catalog, while all other cosmological parameters are fixed to the Planck 2018 best-fit values.

\begin{figure*}[!t]
    \centering
\includegraphics[width=0.7\textwidth]{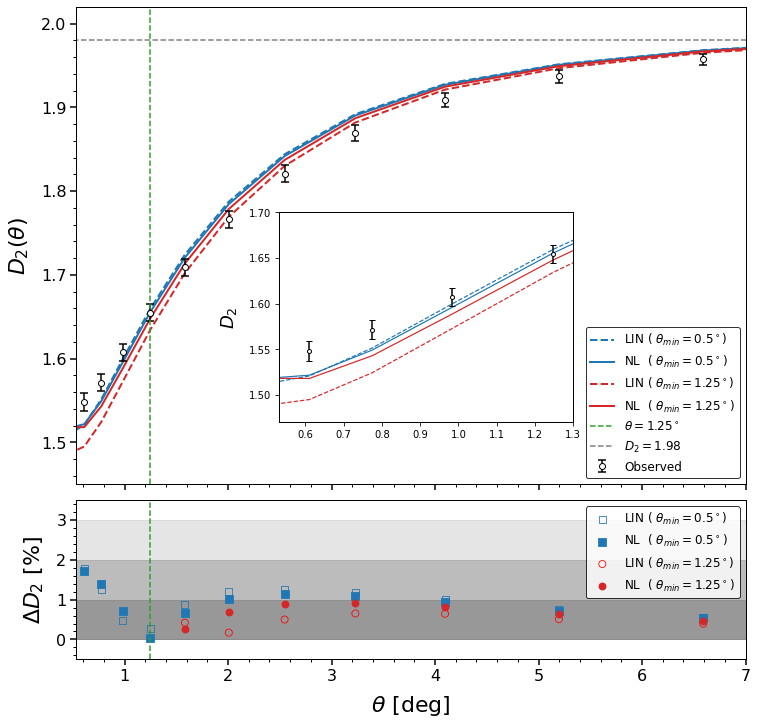} 
    \caption{Top panel: Best-fit values of $ D_{2}(\theta) $ for mock catalogs at $ z = 0.555 $. The dashed and solid blue curves show the results obtained by fitting the mock data with the models defined in Eq.~\ref{ka} and Eq.~\ref{pk2}, respectively, over the scale range $( \theta > 0.5^\circ )$, whereas the dashed and solid red curves correspond to fits using the same models, respectively, but applied over the range $( \theta > 1.25^\circ )$. Black data points with 1-$\sigma$ error bars represent the mock measurements mean values and standard deviations, respectively. The horizontal dashed gray line indicates the homogeneity scale reference value $( D_2 = 1.98 )$, and the vertical dashed green line marks $( \theta = 1.25^\circ )$. An inset zooms in on small angular scales from $ 0.5^\circ$ to  $1.3^\circ $, clearly illustrating the differences between mock measurements and model predictions in that regime.
    Bottom panel: percentage systematic errors on $( D_{2}(\theta) )$, defined as $100 \, ( D_{2}^{model}-D_{2}^{mock} )/D_{2}^{mock}$, i.e., the percentage differences between theoretical predictions and observational data. 
    Symbols indicate residuals for LIN and NL models, with the same color coding as above. The shaded gray bands represent deviations of $\pm 1\%$, $\pm 2\%$, and $\pm 3\%$, illustrating how closely models match observations across different angular scales. Again, the vertical green dashed line again marks the angle $(\theta = 1.25^\circ)$. } 
\label{fig:d2_mock}
\end{figure*}

The top panel of Fig.~\ref{fig:d2_mock} displays the best-fit values of $ D_{2}(\theta) $ as a function of angular separation $\theta $. The dashed and solid curves correspond to fits based on the LIN and nonlinear NL models, respectively. The bottom panel presents the percentage systematic errors, defined as  
$( 100 \cdot ( D_{2}^{\mathrm{model}} - D_{2}^{\mathrm{mock}} ) / D_{2}^{\mathrm{mock}} )$, quantifying the the relative difference between theoretical predictions and mock measurements.

For $\theta_{min} = 0.5^\circ$, with the nonlinear (NL) model (blue solid), we obtain a bias of $b = 1.914 \pm 0.038$ and a pairwise velocity dispersion of $\sigma_p = 263 \pm 40$ km/s, with a reduced $\chi^2 = 0.90$, while, with the linear (LIN) model, we find $b = 1.902 \pm 0.030$ and a reduced $\chi^2 = 0.93$ (blue dashed). These results are consistent with the bias values reported in \cite{Ntelis:2017nrj}, where $b = 1.944 \pm 0.021$ and $\sigma_p = 252 \pm 8.9$ km/s, with a reduced $\chi^2 = 0.91$. The velocity dispersion we find is also in agreement with the typical value for CMASS galaxy samples, $\sigma_p = 240 \pm 50$ km/s \cite{thomas2013stellar}.

As illustrated in the top panel of Fig.~\ref{fig:d2_mock}, when fitting the mock catalog measurements starting from 0.5 degree ($\theta_{min}=0.5^\circ$), both the linear Kaiser model (LIN, blue dashed) and the nonlinear dispersion model (NL, blue solid) provide an good description of \( D_{2}(\theta) \) at angular scales \( \theta > 1.25^\circ \).
However, significant deviations emerge at smaller angular separations (\( \theta < 1.25^\circ \)), where neither model is able to accurately capture the characteristic suppression induced by nonlinear FoG effects. This is anticipated for the LIN model (blue dashed), which neglects nonlinear velocity dispersion. In the case of the NL model (blue solid), the residual discrepancies are attributed to the limitations of the dispersion model in accurately modeling nonlinear RSD, as highlighted in the inset of Fig.~\ref{fig:d2_mock}. While the NL model based on Eq.\ref{pk2} partially mitigates the deviations in \( D_2 \), it fails to fully reconcile the model predictions with the mock measurements — a result consistent with earlier studies (e.g., \cite{mohammad2016group,marulli2017redshift} and references therein).

As shown in the bottom panel of Fig.~\ref{fig:d2_mock}, the best-fit values of \( D_{2}(\theta) \) exhibit systematic deviations from the mock measurements, with the magnitude of the bias varying as a function of angular scale \( \theta \). For fits performed over the full angular range starting from $0.5$ degree ($\theta_{min}=0.5^\circ$), both the linear model (open blue squares) and the nonlinear dispersion model (solid blue squares) yield percentage differences of approximately 2$\%$ at small scales. When the fitting is  performed over the full range starting from $1.25$ degree ($\theta_{min}=1.25^\circ$), the discrepancies are noticeably reduced to  approximately $1\%$ for both the LIN (open red squares) and NL (solid red squares) cases.
Indeed, limiting the full fitting range to start from 1.25 degree  ($\theta_{min} = 1.25^\circ$) hardly affects the results: we obtain $b = 1.951 \pm 0.039$ for the NL model and $b = 1.950 \pm 0.037$ for the LIN model. These values are in excellent agreement with those obtained using $\theta_{min} = 0.5^\circ$, indicating that our bias estimation is quite robust. Furthermore, the derived bias values are consistent with those reported by several other authors \cite{ballinger1996measuring,white2011clustering,nuza2013clustering,ho2012clustering,comparat2013stochastic,hector2015power}.

This result indicates that the angular ranges adopted are sufficiently large to limit the systematic errors on $( D_{2}(\theta) )$. In particular, by selecting a minimum angular scale of $( \theta_{min} = 1.25^\circ )$, one can significantly mitigate systematic biases in $( D_{2}(\theta) )$ without requiring a more sophisticated modeling of nonlinear effects. Notably, discrepancies between observational measurements and $\Lambda$CDM predictions persist, even when employing a range of nonlinear RSD models -- many of which are more accurate than the dispersion model adopted in this work \cite{macaulay2013lower,marulli2017redshift} -- which suggests that residual modeling uncertainties may still be present, even in these more advanced RSD frameworks. In this context, excluding the small-scale regime where nonlinearities become significant appears to be a more effective strategy than attempting to model these effects using the empirical approach of Eq.~\ref{pk2}.

\begin{figure}[!h]
    \centering
\includegraphics[width=0.65  \textwidth]{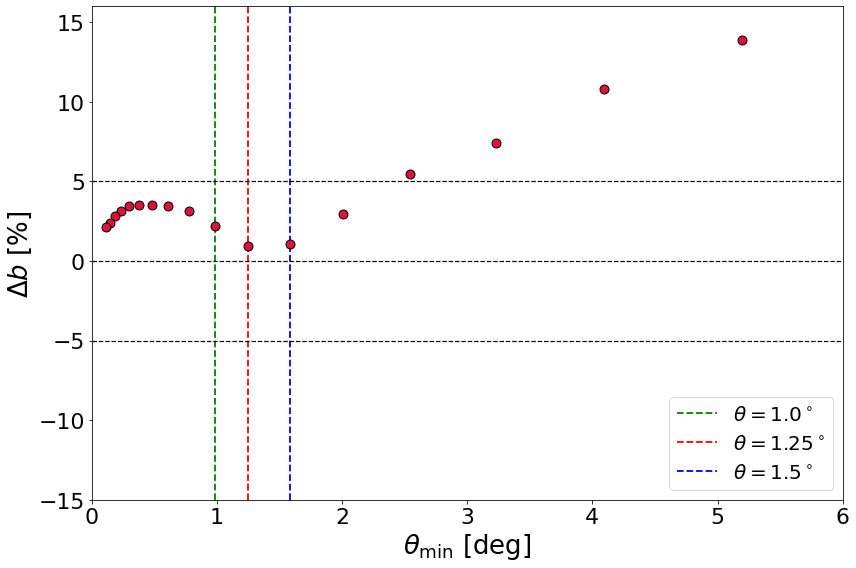} 
    \caption{Relative deviation between the recovered best-fit galaxy bias and the reference bias for the mock catalogs at redshift $z = 0.555$, expressed as $\Delta b = 100 \cdot \frac{b - b_{\rm ref}}{b_{\rm ref}}$. The red points represent the best-fit results obtained for different values of \(\theta_{ min}\), while the vertical dashed lines indicate \(\theta_{min} = 1.0^\circ\) (green), \(\theta_{min} = 1.25^\circ\) (red), and \(\theta_{min} = 1.5^\circ\) (blue).} 
\label{fig:bias_mock}
\end{figure}

We then procedure to apply only the linear Kaiser model (Eq.~\ref{ka}) to estimate the bias parameter using the MultiDark-Patchy mock catalogs (so do not take into account the non-linear FoG effect). The relative difference between the recovered best-fit bias and the reference bias as a function of the minimum angular scale \(\theta_{min}\) is shown in Fig.~\ref{fig:bias_mock}. 
The reference bias ($b_{\mathrm{ref}}$) is obtained from the two-point correlation function analysis described in \cite{Ntelis:2017nrj}.

As shown in Fig.~\ref{fig:bias_mock}, the relative uncertainty of bias increase when $\theta_{min}$ is below 1.25 degrees, where the deviation from the reference bias is larger due to residual FoG effects. However, the uncertainty reaches its minimum
at $\theta_{min}$ = 1.25 degrees. The reduced chi-squared values—$\chi^2_{\nu} = 0.98$ for $\theta_{min} = 1.0^\circ$ and $\chi^2_{\nu} = 0.95$ for $\theta_{min} = 1.25^\circ$—demonstrate that the model without the FoG effect provides a good fit to the data within this scale. For values of $\theta_{min}$ greater than 4 degrees, the relative uncertainty significant rises above $10\%$, primarily due to fewer data points available for analysis.

Our analysis shows that the minimum discrepancies in the bias values (\(\Delta b\)) occur for angular cuts around 
\(\theta_{\min} = 1.0^\circ\) and \(1.25^\circ\).
We have performed an analogous analysis for each individual redshift bin and find that the preferred minimum angular scale is consistently within the range $1.0^\circ$--$1.25^\circ$.
These angles correspond to
physical scales of $18$–$23\,h^{-1}\,\mathrm{Mpc}$ over the redshift range $0.46 \leq z \leq 0.74$. 
For both choices of $\theta_{min}$, the recovered bias remains stable across redshift within the uncertainties, confirming the robustness of the bias recovery.

\newpage

\subsection{Bias measurement on DR12 and DR16 catalogs}\label{model3}

\noindent

As established in the previous section, we adopt a conservative choice of \(\theta_{\min} = 1.25^\circ\) for our subsequent analysis, within which the LIN model—despite neglecting the FoG effect—can reliably recover the galaxy bias parameter. Based on this finding, we now proceed to perform measurements using real data from the SDSS DR12 and DR16 catalogs.

\begin{figure}[!b]
    \centering
\includegraphics[width=0.7\textwidth]{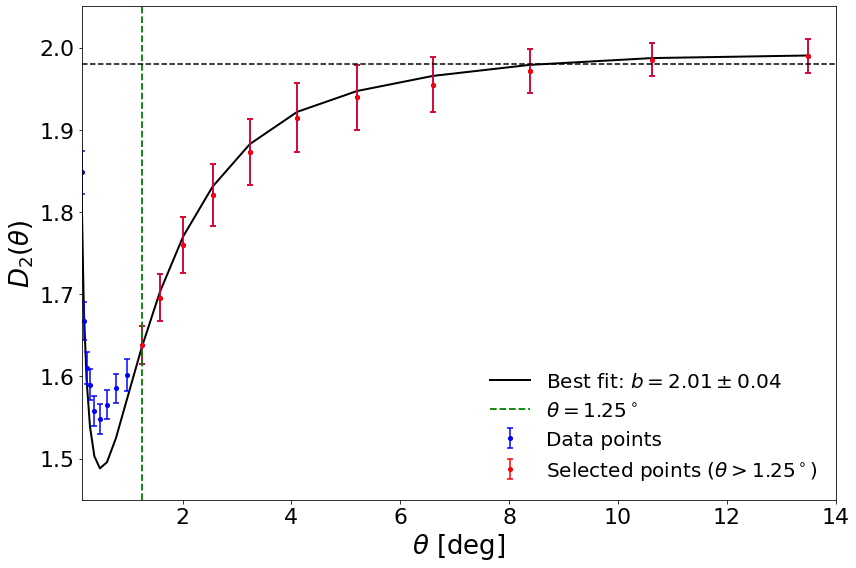} 
    \caption{The correlation dimension \(D_2(\theta)\) as a function of angular separation \(\theta\) at redshift mean $\bar{z}=0.555$ for the real observational data}.
\label{fig:d2_real}
\end{figure}

In Fig.~\ref{fig:d2_real}, we show the correlation dimension \(D_2(\theta)\) as a function of angular separation \(\theta\) at representative redshift \(z=0.555\). The blue points represent all measured data, while the red points indicate the selected subset used for fitting, starting from a minimum angular scale of \(\theta_{min} = 1.25^\circ\) (marked by the green dashed line). The black curve shows the best-fit theoretical model, yielding a bias parameter of \(b = 2.01 \pm 0.04\). Error bars represent the statistical uncertainties in each bin.

\begin{figure}[!t]
    \centering

\includegraphics[width=0.7\textwidth]{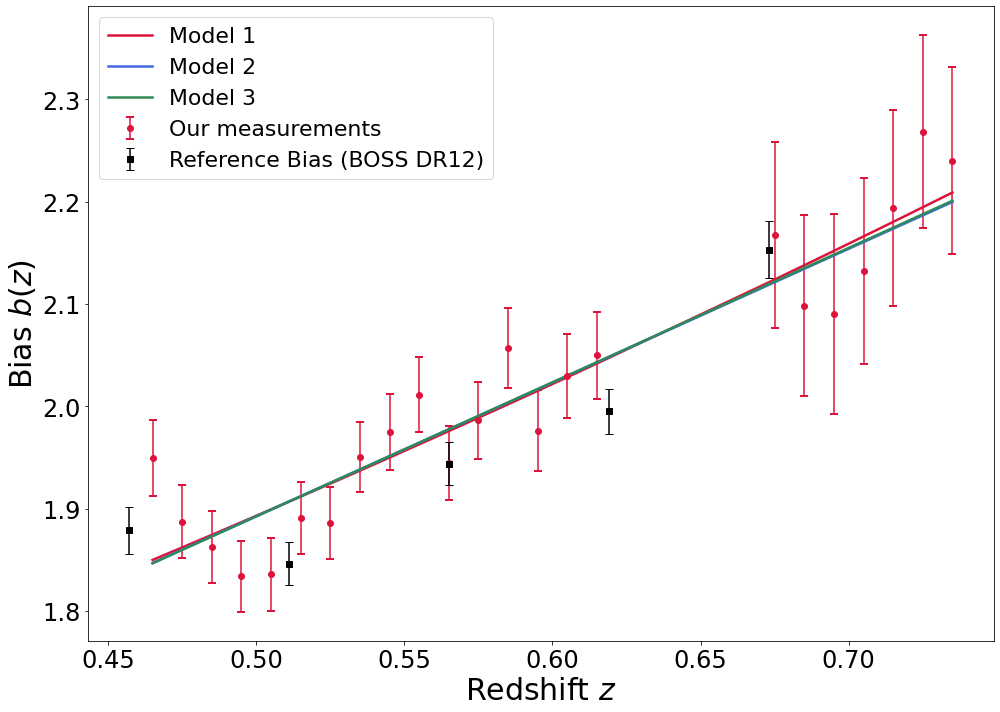} 
    \caption{The measured galaxy bias from the DR12 and DR16 catalogs is shown. The red points represent our measurements, with error bars indicating statistical uncertainties, while the black points correspond to the reference bias values reported in \cite{Ntelis:2017nrj}. The three colored lines represent different bias evolution models: Model 1 (red line), Model 2 (blue line), and Model 3 (green line). } 
\label{fig:bias_compare}
\end{figure} 

We also conduct the bias measurements across the 23 redshift bins spanning the range $ 0.46 \leq z \leq 0.74 $ and presented in Section~\ref{ObsEst}. The full results are shown in Fig.~\ref{fig:bias_compare}, where the measured galaxy bias is shown as a function of redshift using real data. 
The red points with error bars represent our bias measurements, while black points with error bars denote the reference values from Table~2 of \cite{Ntelis:2017nrj}, which are based on the SDSS DR12 analysis, which accounts for the FoG effect.

The figure illustrates the redshift evolution of the galaxy bias parameter $ b(z)$ , along with three theoretical fitting models:
\begin{itemize}
\item Model 1: $ b(z) = a_1 + a_2(1+z)^2 $
\item Model 2: $ b(z) = a_1 + a_2(1+z) $
\item Model 3: $ b(z) = a_1 \frac{1+z}{(1+z_{\text{eff}})^{a_2}} $, where $ z_{\text{eff}} $ denotes the effective redshift of the bin (as in \cite{basilakos2008halo}, \cite{Ntelis:2018ctq}).
\end{itemize}

\begin{table}[h!]
\centering
\begin{tabular}{lccc}
\hline
\textbf{Model} & \( a_1 \) & \( a_2  \) &  \( \chi^2_{\nu} \) \\
\hline
Model 1 & \( 1.064 \pm 0.138 \) & \( 0.37 \pm 0.057 \) & 1.10 \\
Model 2 & \( 0.146 \pm 0.281 \) & \( 1.166 \pm 0.181 \) & 1.09 \\
Model 3 & \( 2.001 \pm 0.013 \) & \( 0.927 \pm 0.142 \) & 1.08 \\
\hline
\end{tabular}
\caption{Best-fit values of \( a_1 \) and \( a_2 \) for the different bias models, along with the corresponding reduced \( \chi^2 \) assuming 21 degrees of freedom.}
\label{bias_model}
\end{table}

In Table~\ref{bias_model} we present the best fit values of the parameters for each model. The corresponding reduced chi-squared values for these fits are 1.1, 1.09, and 1.08, respectively for each model. The Model 3 (shown in green in Fig.~\ref{fig:bias_compare}) provides the best fit, though the improvement over the others is not significant.

Our measurements (red points) show good agreement with the reference values (black points), particularly at intermediate redshifts. At higher redshifts, the scatter and uncertainties increase, due to reduced data density and larger observational errors. As demonstrated above, it is indeed possible to achieve nearly unbiased constraints on the bias factor from $D_{2}$ across all redshifts considered for the DR12 and DR16 samples. A key advantage of this method is the minimal number of free parameters required for the modeling.

\subsection{Bias-Corrected Angular Homogeneity Scale for the Matter Distribution}
\label{b_correct}
\begin{figure}
    \centering
\includegraphics[width=0.49\linewidth]{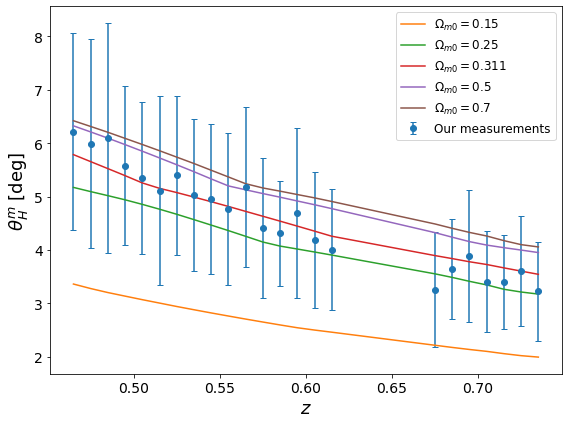}
\includegraphics[width=0.49\linewidth]{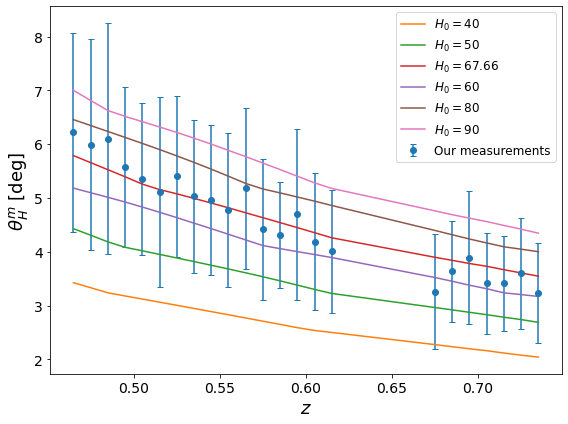}
\caption{
Redshift evolution of the homogeneity angular scale $\theta_H$ for the matter distribution.
Theoretical curves are computed assuming a Planck 2018 $\Lambda$CDM fiducial model.
Left: varying $\Omega_{m0}$ with all other parameters fixed.
Right: varying $H_0$ with the remaining parameters fixed.
Blue points show the bias-corrected angular homogeneity scale $\theta^{m}_{H}$ obtained from the DR12 and DR16 catalogs.}
    \label{fig:thetah_m}
\end{figure}

To compare our results with the SCM prediction, we correct the angular homogeneity index $D_2$ for galaxy bias.
Using the bias measurements obtained from the DR12 and DR16 catalogs (see Fig.~\ref{fig:bias_compare}), we reconstruct the matter angular homogeneity index $D^{m}_2$ according to
\begin{equation}
    D^{m}_2 - 2 = \frac{D^{\mathrm{gal}}_2 - 2}{b^2},
    \label{d2_m}
\end{equation}
where $b$ denotes the linear galaxy bias and $D^{\mathrm{gal}}_2$ is the fractal dimension measured from the galaxy distribution.
From this, we derive the bias-corrected angular homogeneity scale $\theta^{m}_{H}$ (blue points), shown in Fig.~\ref{fig:thetah_m}. 

We compare these measurements with the theoretical prediction computed within a fiducial Planck 2018 $\Lambda$CDM cosmology (red line). We find excellent agreement between the reconstructed $\theta^{m}_{H}$ and the Planck-consistent model, indicating that the observed homogeneity scale is fully compatible with the standard cosmological framework.

The results also confirm the expected increasing trend of the angular homogeneity scale with decreasing redshift, as matter perturbations grow stronger at later epochs and the Universe becomes increasingly clumpy toward lower redshifts. At higher redshifts, the smaller homogeneity scale reflects a smoother Universe in the past, in good agreement with the $\Lambda$CDM prediction over the entire redshift interval analyzed. Our results therefore confirm the presence of a homogeneity scale in the spatial distribution of LRGs, as predicted by the fundamental assumptions of the standard cosmological model.

For comparison, Fig.~\ref{fig:thetah_m} also shows theoretical predictions obtained by varying individual cosmological parameters within the Planck 2018 $\Lambda$CDM framework. 
In the left panel, $\Omega_m$ is varied while keeping all other parameters fixed; 
in the right panel, $H_0$ is varied with the remaining parameters fixed. 
We find that the angular homogeneity scale $\theta^{m}_{H}$ is sensitive to both  $H_0$ and $\Omega_{m0}$ (with $\Omega_{m0} = 1 - \Omega_{\Lambda}$ in a flat universe), which is expected since an accelerating expansion suppresses the growth of structures, thereby decreasing the homogeneity scale~\cite{Ntelis:2018ctq}. 

Therefore, it would be interesting to explore constraining the cosmological parameters $\Omega_{m0}$ and $H_0$ using the homogeneity scale. Indeed, several studies have already derived cosmological constraints from the homogeneity scale \cite{Ntelis:2018ctq,Shao:2024qrd} and references therein.

In the next Section, our main objective is to extend these previous analyses by considering a broader range of the angular homogeneity index $D_2$. Specifically, we investigate how the  angular homogeneity index  $D_2(\theta)$ can be used to constrain cosmological parameters $\omega_m = \Omega_{m0}h^2$. In particular, we demonstrate that adopting an appropriate minimum angular scale cut when fitting the $D_2(\theta)$ curve effectively suppresses the FoG contribution, thereby rendering the residual nonlinear effects negligible.

This strategy requires only a minimal set of free parameters, simplifying the modelling while maintaining robustness. We emphasize that our extraction of $D_2(\theta)$ from observational data does not assume any fiducial cosmology. However, in order to perform parameter constraints, we adopt the standard $\Lambda$CDM model framework.

\newpage
\section{Cosmological Estimation}
\label{sec:result}
From hereafter we perform a cosmological inference of the physical matter density, $\omega_m$, based on the angular homogeneity index $D_2(\theta)$. We begin by validating our methodology for selecting the minimum angular scale cut, $\theta_{\min}$, using mock catalogs. In Section~\ref{sec:result.mock}, this validation is performed using mock catalogs in order to verify that the method can recover the true input cosmology assumed in the mock catalogs. After, in Section~\ref{sec:result.data} we then apply the validated procedure to the real observational data.

\subsection{Validation with Mock Catalogues}
\label{sec:result.mock}

We use mock catalogs from the MultiDark-Patchy mock catalogs \cite{kitaura2016clustering, rodriguez2016clustering} and the EZmock galaxy catalogues \cite{zhao2021completed} and perform a Markov Chain Monte Carlo (MCMC) analysis, based on measurements of the correlation dimension \( D_{2} \). The analysis is made within the framework of a spatially flat \(\Lambda\)CDM model, 
where the physical matter density $\omega_m = \Omega_m h^2$ is derived from the varying values of the free parameters $\Omega_m$ and $h$.
The bias $b$ is modeled using Bias Model 3 (Section~\ref{model3}) and we marginalize over the nuisance bias parameters $a_1$ and $a_2$. We have also tested alternative bias models and found that they do not significantly affect our results. The remaining fiducial cosmological parameters, \( p_F = (n_s, \ln[10^{10} A_s], \Omega_k) \), are fixed to their best-fit values from the Planck 2018 results \cite{aghanim2021planck}. The chi-squared function used in this analysis is defined as:
\begin{eqnarray}
\chi^{2} = \sum_{ij} 
\left[ D_{2}^{\mathrm{data}} - D_{2}^{\mathrm{th}}(b, \omega_m, p_{F}) \right] 
\, C^{-1}_{ij} \,
\left[ D_{2}^{\mathrm{data}} - D_{2}^{\mathrm{th}}(b, \omega_m, p_{F}) \right]
\label{eq:chi_cov_sum}
\end{eqnarray}
where \( D_{2}^{\mathrm{data}} \) denotes the correlation dimension measured from the data, and \( C_{ij} \) is the covariance matrix derived from mock catalogues. 
The theoretical prediction \( D_{2}^{\mathrm{th}} \) represents the model correlation dimension of the galaxy distribution corresponding to the theoretical curve obtained from the parameters previously described.
To account for the finite number of mock realizations, we correct the inverse covariance matrix using the Hartlap factor \cite{ taylor2013putting},
\begin{equation}
\Psi_{ij} = \frac{N_{\mathrm{mock}} - N_{\mathrm{bin}} - 2}{N_{\mathrm{mock}} - 1} \, C^{-1}_{ij},
\end{equation}
where \(N_{\mathrm{mock}}\) is the number of mock realizations and \(N_{\mathrm{bin}}\) is the number of data bins.

In addition to rescaling the inverse covariance matrix $\psi_{ij}$, we further account for the propagation of covariance matrix noise into the uncertainties of the inferred parameters following \cite{percival2014clustering}. 
To correct the parameter uncertainties when fitting the mocks catalogss, we apply the  correction factor $m_{2}$, given by
\begin{equation}
m_{2}=m_{1} \sqrt{\frac{N_{\mathrm{mock}}-1}{N_{\mathrm{mock}}-N_{\mathrm{bin}}-2}} .
\end{equation}
Here, $m_1$ is the correction factor appropriate for parameter uncertainties inferred from fits to independent observational data, and is given by
\begin{equation}
\label{m1}
m_{1}=\frac{1+B\left(N_{\mathrm{bin}}-N_{\mathrm{p}}\right)}{1+A+B\left(N_{\mathrm{p}}+1\right)} ,
\end{equation}
where $N_p$ is the number of fitted parameters, and
\begin{equation}
A=\frac{2}{\left(N_{\mathrm{mock}}-N_{\mathrm{bin}}-1\right)\left(N_{\mathrm{mock}}-N_{\mathrm{bin}}-4\right)} ,
\end{equation}
\begin{equation}
B=\frac{N_{\mathrm{mock}}-N_{\mathrm{bin}}-2}{\left(N_{\mathrm{mock}}-N_{\mathrm{bin}}-1\right)\left(N_{\mathrm{mock}}-N_{\mathrm{bin}}-4\right)} .
\end{equation}
With $N_{\mathrm{mock}} = 1000$ and $N_{\mathrm{bin}} = (23\text{--}15)\times 23$, depending on the choice of $\theta_{\min}$, both $m_2$ and $m_1$ correction factors are substantial and need to be accounted for when fitting both mock and real catalogues.

\begin{figure*}[!h]
    \centering

\includegraphics[width=0.7\textwidth]{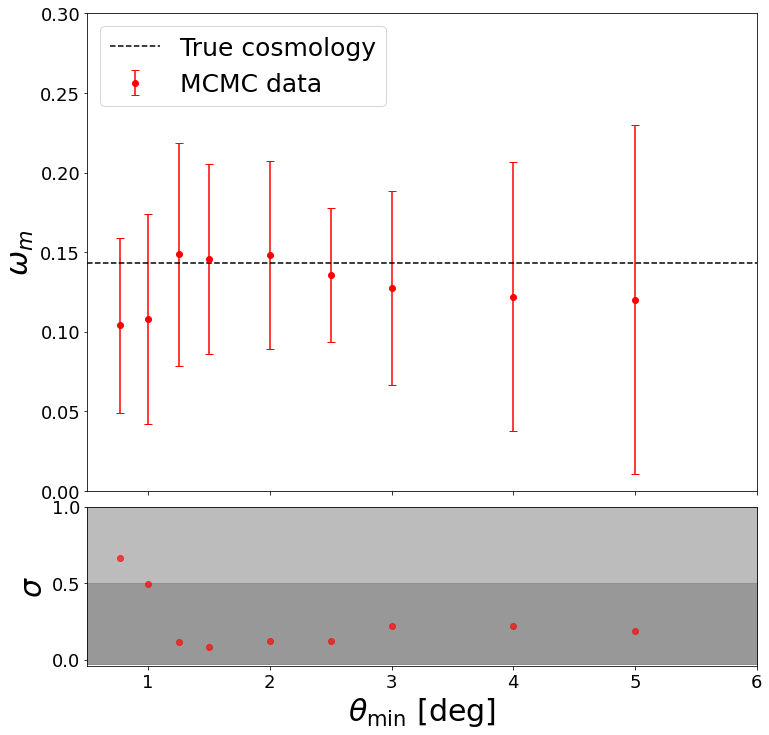} 
    \caption{Cosmological constraints on $\omega_m$ from mock catalogs and their  comparison with the true input value assumed in the mock catalogues at $1\sigma$ confidence level as a function of $\theta_{min}$. The shaded horizontal bands correspond to $0.5\sigma$ (dark gray), $1\sigma$ (gray) calculated by Eq.~\ref{Tension}.} 
\label{fig:sig_mock}
\end{figure*}

Figure~\ref{fig:sig_mock} shows the constraints on the physical matter density \(\omega_m\) and its comparison with the true input value assumed in the mock catalogues, as a function of the minimum angular scale \(\theta_{min}\). The top panel shows that the best-fit \(\omega_m\) values obtained from MCMC fits to the correlation dimension \(D_2\), with red points and error bars representing the measurements and their uncertainties. 
The solid lines indicate the true input value of \(\omega_m\) assumed in the mock catalogues.

Moreover in order to quantify the agreement between the inferred $\omega_m$ from the LIN model and the true input value, we define a tension metric:
\begin{equation}
\label{Tension}
T_1 = \frac{|\mu_{\mathrm{D}_2} - \mu_{\mathrm{true}}|}{\sigma_{\mathrm{D}_2} },
\end{equation}
where $\mu_{\mathrm{D}_2}$ and $\sigma_{\mathrm{D}_2}$ are the mean and uncertainty of $\omega_m$ from our analysis, and $\mu_{\mathrm{true}}$ is the true input value. We evaluate this tension as a function of angular scale to understand the scale-dependent influence of non-linearities on cosmological parameter estimation.

The bottom panel of Figure~\ref{fig:sig_mock} is obtained from Eq.~\ref{Tension} and it shows the corresponding tension in units of standard deviation \(\sigma\) relative to the true input value. Shaded regions mark the \(0.5\sigma\), \(1\sigma\) confidence levels. For the sake of example, for $\theta_{min} = 1.25^o$ we obtain a constraint of \(\omega_{m} (\theta_{min} = 1.25^o) = 0.149^{+0.058}_{-0.079}\) from mock catalogs, which is in excellent agreement with the true input value  with a discrepancy of  \(0.1\sigma\).\footnote{
As a consistency check, we also consider a fixed minimum physical scale, $r_{\min} = 20\,h^{-1}\,\mathrm{Mpc}$, which is mapped to a redshift-dependent angular scale $\theta_{\min}(z)$ using the fiducial cosmology. This redshift-dependent cut is then applied to each redshift bin. 
We obtain $\omega_m = 0.145^{+0.044}_{-0.072}$ from the mock data, in agreement with the result derived using a fixed angular cut of $\theta_{\min} = 1.25^\circ$.
}
The corresponding best-fit cosmological parameters in this case are $H_0 = 72.7 \pm 17.1 ~\rm{km \cdot s^{-1}\cdot Mpc^{-1}}$ and $\Omega_m = 0.279 \pm 0.064$.

As shown in Figure~\ref{fig:sig_mock}, for low values of $\theta_{\min}$ (e.g., below $1^\circ$), the tension reaches values of up to $\sim 1\sigma$, indicating a mild deviation between the $\omega_m$ inferred from $D_2$ and the  true input value. This deviation is primarily driven by non-linear FoG effects, which distort clustering measurements on small angular scales due to random peculiar velocities within collapsed structures.

As $\theta_{\min}$ increases to values between $1.25^\circ$ and $3^\circ$, the tension drops below $0.5\sigma$, and the constraint on \(\omega_m\) becomes stable, indicating significantly improved consistency between the $D_2$-based inference and the true input value. For $\theta_{\min} > 3^\circ$, although the tension remains below $0.5\sigma$, the uncertainty on the inferred $\omega_m$ increases significantly due to the reduced number of available data points at larger angular scales.

These results suggest that angular scales in the range $\theta_{\min} \simeq 1.25^\circ$–$3^\circ$ provide an optimal compromise between minimizing tension and avoiding non-linear distortions, making them the most suitable for robust cosmological parameter estimation from $D_2$.

\subsection{Cosmological Constraints from Real data}
\label{sec:result.data}

\begin{figure*}[!t]
    \centering
\includegraphics[width=0.7\textwidth]{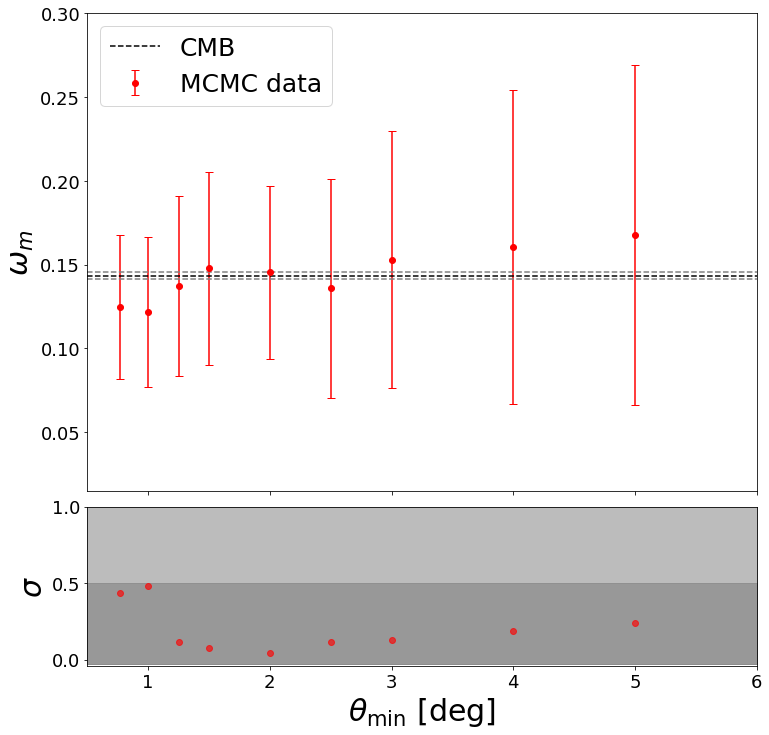} 
    \caption{Cosmological constraints on $\omega_m$ from real catalogs and their comparison with CMB at $1\sigma$ confidence level as a function of $\theta_{min}$. The shaded horizontal bands correspond to $0.5\sigma$ (dark gray), $1\sigma$ (gray) intervals  calculated by Eq.~\ref{Tension2}.} 
\label{fig:sig_real}
\end{figure*}

After validating the methodology with mock catalogs, we apply the same analysis to the real galaxy clustering data from SDSS DR12 and DR16. Figure~\ref{fig:sig_real} shows the inferred $\omega_m$ and its comparison with  the CMB value as a function of $\theta_{\min}$ from the $D_2$ analysis of the real data. The dashed lines indicate the Planck 2018 CMB constraint on \(\omega_m\), corresponding to the $1\sigma$ uncertainty.

To quantify the agreement between the value of $\omega_m$ inferred from the LIN model and the Planck 2018 result, we define the following tension metric:
\begin{equation}
\label{Tension2}
T_2 = \frac{|\mu_{\mathrm{D}_2} - \mu_{\mathrm{Planck}}|}{\sqrt{\sigma^2_{\mathrm{D}_2} + \sigma^2_{\mathrm{Planck}}}},
\end{equation}
where $\mu_{\mathrm{Planck}}$ and $\sigma_{\mathrm{Planck}}$ denote the corresponding mean value and uncertainty from the \textit{Planck} 2018 results. 
The uncertainty on $\omega_m$ inferred from our analysis of the observational data, $\sigma_{\mathrm{D}_2}$, is corrected by the factor $m_1$ (see Eq.~\ref{m1}).

The results from real data analysis are used here as a consistency check of the trend identified in the mock-based validation, rather than as an independent criterion to define a reliable $\theta_{\min}$ interval. At $\theta_{\min}=1.25^\circ$, we obtain a best-fit value of $\omega_m = 0.137^{+0.041}_{-0.059}$, which is in excellent agreement with the \textit{Planck} 2018 result, corresponding to a tension of only $0.1\sigma$. We also observe the similar trend found in the mock analysis: as $\theta_{\min}$ increases from $1.25^\circ$ to $3^\circ$, the inferred values of $\omega_m$ become stable and remain within $0.5\sigma$ of the CMB constraint, indicating reduced sensitivity to small-scale nonlinearities. 

Motivated by the mock validation, we therefore adopt $\theta_{\min}=1.25^\circ$ as our fiducial minimum angular scale and constrain the cosmological parameters $\Omega_{m0}$ and $H_0$. The prior on $\Omega_{m0}$ is taken to be uniform in the range $[0.05,1.0]$, and the prior on $H_0$ is uniform in the range $[20,100]$. We obtain a best-fit Hubble constant of $H_0 = 71.6 \pm 14.0~\rm{km \cdot s^{-1}\cdot Mpc^{-1}}$, and a best-fit matter density parameter of \(\Omega_{m0} = 0.266 \pm 0.055\). The result is in good agreement with the independent analysis of \cite{holanda2019estimate}, which used gas mass fraction measurements in galaxy clusters and obtained \(\Omega_{m0} = 0.285 \pm 0.013\), corresponding to a discrepancy of only \(0.34\sigma\). Our best-fit value of \(\Omega_{m0}\) is also in agreement with the result of \cite{semenaite2111cosmological}, within \(0.36\sigma\), where they reported \(\Omega_{m0} = 0.286^{+0.011}_{-0.013}\). Their analysis is based on the average of the two-point correlation function  using combined data from BOSS, eBOSS, and DES, and includes information from the range (\(20\,h^{-1}\,\mathrm{Mpc} < s < 160\,h^{-1}\,\mathrm{Mpc}\)). For our analysis, we restrict the minimum angular separation to \(\theta_{min} = 1.25^\circ\), corresponding to 
physical scales of $18$–$23\,h^{-1}\,\mathrm{Mpc}$ over the redshift range $0.46 \leq z \leq 0.74$, with the fiducial model used. We note that our best-fit value of \(\Omega_{m0}\) is slightly smaller (but still consistent within \(1\sigma\)) compared to the value derived from BAO datasets, which combine measurements from the SDSS DR7 main galaxy sample \cite{ross2015clustering}, the 6dF Galaxy Survey \cite{beutler20116df}, and the SDSS BOSS DR12 \cite{alam2017clustering} and eBOSS DR16 surveys \cite{alam2021completed}, without assuming priors on the sound horizon scale at the drag epoch \(r_{\rm drag}\), yielding \(\Omega_{m0} = 0.293^{+0.13}_{-0.15}\). Nonetheless, this difference is not unexpected, given the distinct methodologies employed.

\section{Conclusions}\label{sec:conclu}

\noindent

In this work, we addressed the challenge posed by the inaccurate modeling of non-linear motions at small scales within the dispersion model, which can lead to biased constraints on the galaxy bias factor and the physical matter density \(\omega_m\). Rather than attempting to refine the model to better capture small-scale nonlinearities, we demonstrate that nearly unbiased constraints on the bias factor and cosmological parameters can be obtained by modeling the angular correlation dimension function $D_2(\theta)$. This approach is valid since the analysis is restricted to sufficiently large scales where non-linear effects are negligible, then neglecting the necessity of robustly modeling the uncertainties and improving RSD at small scales.

The angular correlation dimension $D_2(\theta)$ is a cumulative statistic derived from the two-point correlation function and serves as a powerful and model-independent probe of cosmic homogeneity. Initially we found the cosmic homogeneity scale $\theta_{H}$ values for 23 bins across the redshift range $0.46 < z < 0.74$, based on the observational SDSS DR12 and DR16 Luminous Red Galaxy (LRG) catalogs. We then comprehensively explored the full angular interval of the $D_2$ curve, so differently from the general literature of cosmic homogeneity that basically focus on the $\theta_{H}$. We developed such a study by varying the starting angles $\theta_{min}$, then we systematically defined and investigated the impact of $\theta_{min}$ on our analyses and estimates of cosmological parameters.

The next step was to investigate the RSD effects, namely the Kaiser effect (a linear modeling that depends on the bias factor) and the non-linear FoG effect (dependent on the velocity dispersion). We performed a theoretical analysis demonstrating how the theoretical $D_2$ curves change based on the peculiar velocity assumed in the non-linear model. To validate our methodology, we fitted realistic mock catalogs using only the linear model within a $\Lambda$CDM fiducial model, with bias as the only free parameter, testing various $\theta_{min}$ values. 
This analysis reveals how $\theta_{\min}$ impacts the fitted bias and identifies the optimal choice of angular scales, which correspond to sufficiently large comoving distances, allowing the use of the simpler linear model and avoiding to characterize the non-linear model, which demands a precise understanding of the assumed velocity dispersion.

Based once more on the LRG catalogs from SDSS DR12 and DR16 we calculated the linear bias factor for the same 23 redshift bins previously mentioned. We obtained bias values typically between $1.8$ and $2.3$, consistent with a redshift evolution. We then fitted three different theoretical parameterizations to the observed bias evolution in redshift, finding no significant discrepancies between them.

Finally, we validated our methodology using mock catalogs to ensure that the inferred cosmological parameter (the physical matter density $\omega_m$) is consistent with the input cosmology.
This validation allowed us to identify an optimal range for the minimum angular scale  emerging as the smallest scale that yields results in agreement with the \textit{Planck} CMB constraints.
With this method validated, we proceeded to apply the same methodology to real observational data, within the context of the $\Lambda$CDM model. From the $D_2$ analysis using $\theta_{\min} = 1.25^\circ$, we obtained a best-fit value of $\omega_m = 0.137^{+0.028}_{-0.042}$, which is in excellent agreement with the \textit{Planck} 2018 result.
Our results demonstrate that the angular correlation dimension $D_2(\theta)$
provides a powerful and complementary probe of cosmology, offering robust and unbiased constraints on cosmological parameters by judiciously handling RSD at small scales.

\acknowledgments

XS is supported by Funda\c{c}\~ao de Amparo \`a Pesquisa do Estado do Rio de Janeiro (FAPERJ). RSG thanks financial support from FAPERJ grant No. 260003/005977/2024 - APQ1. CB acknowledges financial support from the CNPq grant 306630/2025-7. JSA is supported by CNPq grant No. 307683/2022-2 and FAPERJ grant No. 259610 (2021). This work was developed thanks to the use of the National Observatory Data Center (CPDON).










\bibliographystyle{unsrt}
\bibliography{main_v1}

\newpage

\begin{appendices}

\section{Comparison of Bootstrap and Mock-derived Uncertainties}

We present a comparison between bootstrap-based uncertainties and those derived from mock realizations. The same bootstrap procedure used for the data catalogue is applied to three representative mock realizations (realizations 0, 3, and 5), and the resulting uncertainties are compared with those obtained from the full ensemble of 1000 EZmock realizations.

\begin{figure}[!h]
    \centering

\includegraphics[width=0.7\textwidth]{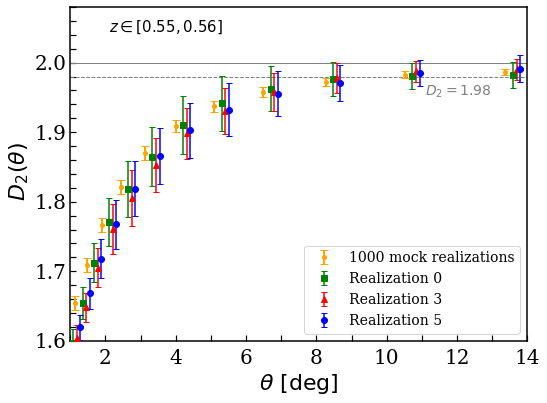} 
\caption{$D_{2}(\theta)$ measured in the redshift slice $z\in[0.55,0.56]$. The dashed line marks the large-scale asymptote $D_{2}=1.98$.
The orange data points indicate the $1\sigma$ scatter derived from the ensemble of 1000 mock realizations. 
Colored markers (green, red, and blue) correspond to three representative mock realizations (0, 3, and 5), with their bootstrap uncertainties. }

\label{fig:realization}
\end{figure}

As shown in Fig.~\ref{fig:realization}, the bootstrap uncertainties estimated from individual mocks are systematically larger than those inferred from the mock ensemble, indicating that the bootstrap method overestimates the statistical uncertainties compared to those derived from the ensemble of mock realizations.

\vspace{2cm}

\newpage

\section{Angular Homogeneity Scale}
\label{ThetaHBinsZ}

We present the angular homogeneity scale $\theta_H$ measured across all redshift bins considered in our analysis. Table.\ref{theta_all} summarizes the values of $\theta_H$ obtained using two independent approaches: bootstrap resampling of the data and measurements from mock catalogues. The bootstrap results, $\theta_H^{\mathrm{boot}}$, reflect the statistical uncertainties derived directly from the observational sample, while $\theta_H^{\mathrm{mock}}$ provides an estimate based on mock realizations. 

\begin{table}[h!]
\centering
\begin{tabular}{ccc}\hline $\overline{\mathbf{z}}$ & \(\theta_{H}^{\text {boot }}[\mathrm{deg}]\) & \(\theta_{H}^{\text {mock }}[\mathrm{deg}]\) \\ \hline 0.465 & \(10.4 \pm 2.47\) & \(12.07 \pm 1.46\) \\ 0.475 & \(10.43 \pm 2.58\) & \(11.70 \pm 1.53\) \\ 0.485 & \(9.70 \pm 2.73\) & \(11.44 \pm 1.53\) \\ 0.495 & \(9.08 \pm 2.50\) & \(11.30 \pm 1.55\) \\ 0.505 & \(9.14 \pm 2.48\) & \(11.18 \pm 1.57\) \\ 0.515 & \(9.28 \pm 2.76\) & \(10.97 \pm 1.55\) \\ 0.525 & \(9.57 \pm 2.39\) & \(10.71 \pm 1.56\) \\ 0.535 & \(9.80 \pm 2.67\) & \(10.63 \pm 1.49\) \\ 0.545 & \(9.72 \pm 2.69\) & \(10.36 \pm 1.54\) \\ 0.555 & \(9.17 \pm 2.54\) & \(10.17 \pm 1.48\) \\ 0.565 & \(9.35 \pm 2.77\) & \(10.20 \pm 1.51\) \\ 0.575 & \(9.07 \pm 2.75\) & \(10.08 \pm 1.47\) \\ 0.585 & \(8.31 \pm 2.63\) & \(9.78 \pm 1.43\) \\ 0.595 & \(8.88 \pm 2.87\) & \(9.63 \pm 1.38\) \\ 0.605 & \(8.54 \pm 2.90\) & \(9.49 \pm 1.35\) \\ 0.615 & \(8.36 \pm 2.90\) & \(9.32 \pm 1.24\) \\ 0.675 & \(7.49 \pm 2.83\) & \(8.53 \pm 1.73\) \\ 0.685 & \(8.53 \pm 2.58\) & \(8.51 \pm 1.84\) \\ 0.695 & \(8.07 \pm 2.59\) & \(8.41 \pm 1.74\) \\ 0.705 & \(7.36 \pm 2.49\) & \(8.09 \pm 1.70\) \\ 0.715 & \(6.92 \pm 2.28\) & \(7.93 \pm 1.70\) \\ 0.725 & \(8.12 \pm 2.04\) & \(8.01 \pm 1.74\) \\ 0.735 & \(8.89 \pm 2.90\) & \(7.87 \pm 1.66\) \\ \hline\end{tabular}
\caption{Angular homogeneity scales \(\theta_H\) obtained from both bootstrap resampling and mock catalogs, across all redshift bins. 
}
\label{theta_all}
\end{table}

\section{Bin width analysis}
\label{bin7}

In our analysis, the signal-to-noise ratio is defined as
\begin{equation}
\mathrm{SNR}(\theta) = \frac{2 - D_2(\theta)}{\sigma_{D_2}(\theta)},
\end{equation}
which quantifies the departure from cosmic homogeneity at a given angular scale $\theta$. 

\begin{figure}[!h]
    \centering
\includegraphics[width=0.7\textwidth]{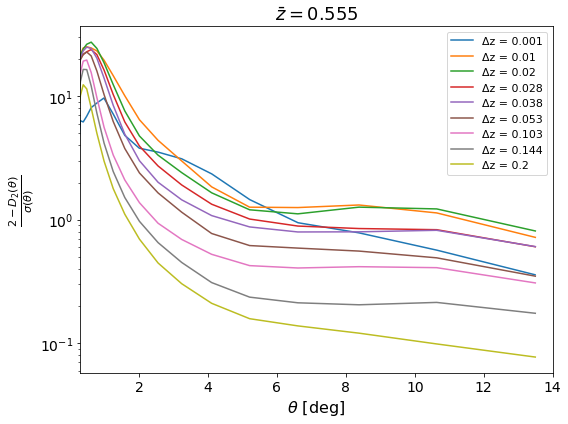} 
\caption{Departure from homogeneity, quantified as 2-$D_2 / \sigma_{D_2}$, as a function of $\theta$ for a redshift bin centered at $\bar{z} = 0.555$ with different bin widths.}
\label{fig:dep_d2}
\end{figure}

As shown in Fig.~\ref{fig:dep_d2}, we compare this quantity for a redshift bin centered at $\bar{z} = 0.55$ using several values of $\Delta z$. We find that increasing the redshift bin width (e.g., $\Delta z = 0.2$) leads to stronger projection effects, which make the observed galaxy distribution appear more homogeneous—that is, 2-$D_2(\theta)$ decrease. Although the uncertainty $\sigma_{D_2}$ also decreases due to the higher number of galaxies in wider bins, the decrease in 2-$D_2(\theta)$ is more pronounced than the reduction in its uncertainty, resulting in an overall lower signal-to-noise ratio.

Conversely, very narrow bins (e.g., $\Delta z = 0.001$) lead to high statistical noise, reducing the signal-to-noise ratio at both small and large angular scales due to limited galaxy counts.

We have verified that this behavior holds consistently across all redshift in our analysis. Taking into account  the signal-to-noise ratio, we adopt $\Delta z = 0.01$. It avoids excessive projection effects while maintaining sufficiently high signal-to-noise across all angular scales.

\begin{figure}[!htbp]
    \centering
\includegraphics[width=0.9\textwidth]{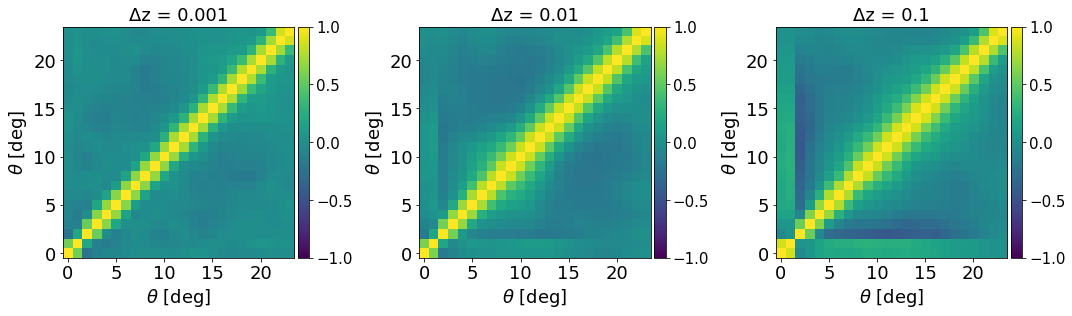} 
\caption{
Correlation matrices of $D_2(\theta)$ measured from 1000 mock realisations, shown for three different redshift bin widths: $\Delta z = 0.001$, $0.01$, and $0.1$ at redshift $\bar{z}=0.555$ .
}
\label{fig:cov_width}
\end{figure}

Figure~\ref{fig:cov_width} further illustrates how redshift bin width impacts the correlation structure of $D_2(\theta)$. The panels show the correlation matrix $\rho_{i,j} = C_{i,j} / \sqrt{C_{i,i} C_{j,j}}$ at $\bar{z} = 0.555$ for three different bin widths: $\Delta z = 0.001$, $0.01$, and $0.1$. As the bin width increases, off-diagonal correlations become more significant due to stronger projection effects. In particular, when using $\Delta z = 0.01$, the off-diagonal correlations are non-negligible.

\begin{figure}[!htbp]
    \centering
\includegraphics[width=0.33\textwidth]{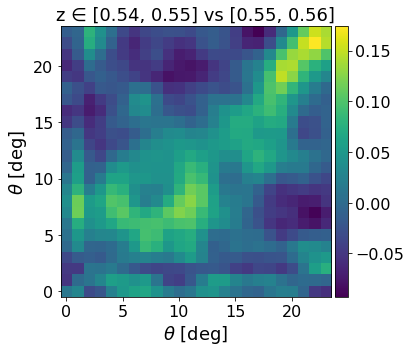} 
\includegraphics[width=0.33\textwidth]{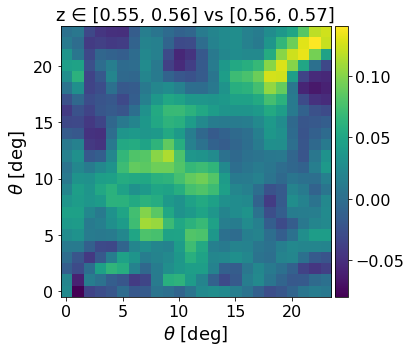} 
\caption{Correlation matrices of $D_2(\theta)$ for cross-covariances between neighbouring redshift bins: $[0.54,0.55]$–$[0.55,0.56]$ (left) and $[0.55,0.56]$–$[0.56,0.57]$ (right).}
\label{fig:cov_cross}
\end{figure}

We also examine the cross-correlations between neighbouring redshift bins for $\Delta z = 0.01$, as shown in Figure~\ref{fig:cov_cross}. The correlation matrices correspond to the redshift bins $[0.54,0.55]$–$[0.55,0.56]$ (left) and $[0.55,0.56]$–$[0.56,0.57]$ (right). A non-negligible level of cross-correlation is observed between these bins, indicating that the measurements are not fully independent. Therefore, the full covariance matrix, including both angular correlations and correlations between redshift bins, should be taken into account in the analysis.

\end{appendices}

\end{document}